\documentclass[11pt,a4paper]{article}
\usepackage{jcappub} 
\usepackage{bm}
\usepackage[compat=1.1.0]{tikz-feynhand}
\usepackage{feynmp}
\usepackage{amsmath}
\usepackage{mathrsfs}
\usepackage[mathscr]{euscript}
\usepackage{feynmp}
\usepackage{comment}
\usepackage{natbib}
\usepackage{ulem}

\allowdisplaybreaks[3]

\def\Mpl{M_{\rm Pl}}

\def\0{{(0)}}
\def\sig0{\dot{\sigma}_0}

\def\ph0{\dot{\phi}_0}

\usepackage{scalerel,stackengine}
\stackMath
\newcommand\reallywidehat[1]{%
\savestack{\tmpbox}{\stretchto{%
  \scaleto{%
    \scalerel*[\widthof{\ensuremath{#1}}]{\kern-.6pt\bigwedge\kern-.6pt}%
    {\rule[-\textheight/2]{1ex}{\textheight}}
  }{\textheight}%
}{0.5ex}}%
\stackon[1pt]{#1}{\tmpbox}%
}
\title{
  Parity-violating scalar trispectrum from a rolling axion during inflation
}

\author[a,b]{Tomohiro Fujita,}
\author[c]{Tomoaki Murata,}
\author[d]{Ippei Obata,}
\author[e]{Maresuke Shiraishi}

\affiliation[a]{Waseda Institute for Advanced Study, Waseda University, Shinjuku, Tokyo 169-8050, Japan}
\affiliation[b]{Research Center for the Early Universe (RESCEU), University of Tokyo, Bunkyo, Tokyo 113-0033, Japan}
\affiliation[c]{Department of Physics, Rikkyo University, Toshima, Tokyo 171-8501, Japan}
\affiliation[d]{Kavli Institute for the Physics and Mathematics of the Universe (Kavli IPMU, WPI),UTIAS, The University of Tokyo, 5-1-5 Kashiwanoha, Kashiwa, Chiba, 277-8583, Japan}
\affiliation[e]{School of General and Management Studies, Suwa University of Science, Chino, Nagano 391-0292, Japan}

\emailAdd{tomofuji@aoni.waseda.jp}
\emailAdd{tmurata@rikkyo.ac.jp}
\emailAdd{ippei.obata@ipmu.jp}
\emailAdd{shiraishi\_maresuke@rs.sus.ac.jp}

\abstract{
We study a mechanism of generating the trispectrum (4-point correlation) of curvature perturbation through the dynamics of a spectator axion field and U(1) gauge field during inflation. 
Owing to the Chern-Simons coupling, only one helicity mode of gauge field experiences a tachyonic instability and sources scalar perturbations. 
Sourced curvature perturbation exhibits parity-violating nature which can be tested through its trispectrum.
We numerically compute parity-even and parity-odd component of the sourced trispectrum. 
It is found that the ratio of parity-odd to parity-even
mode can reach $\mathcal{O}(10\%)$ 
in an exact equilateral momentum configuration.
We also investigate 
a quasi-equilateral shape where only one of the momenta is slightly longer than the other three, and find that the parity-odd mode can reach, and more interestingly, surpass the parity-even one. This may help us to interpret a large parity-odd trispectrum signal extracted from BOSS galaxy-clustering data.
}

\begin{document}

\null\hfill\begin{tabular}[t]{l@{}}
  { RUP-23-20}
\end{tabular}

\maketitle

\section{Introduction}

Is the parity symmetry of our universe broken?
While the standard model of elementary particles (SM) violates the parity symmetry, its effect would be negligible for physics on large scales in our universe.
Also, general relativity (GR), which governs the time evolution of our universe, respects the parity symmetry. 
Hence, exploring the cosmological parity-violating phenomena may help us to search for new physics beyond SM or GR, such as identifying the nature of the dark sector described in the standard $\Lambda$ cold dark matter ($\Lambda$CDM) scenario.

Temperature and E/B-mode polarization of the cosmic microwave background (CMB) are 
powerful observables of cosmological parity violation. A cross correlation between the temperature/E-mode polarization field and the B-mode polarization one, i.e., a $TB$/$EB$ correlation, that is parity-odd quantity, is
a frequently-used diagonostic tool \cite{Lue:1998mq}. Refs.~\cite{Minami:2020odp,Diego-Palazuelos:2022dsq,Eskilt:2022wav,Eskilt:2022cff,Eskilt:2023ndm} have found $> 3\sigma$ signals in the $EB$ correlation from the {\it Planck} and WMAP data. These are likely to be produced by rotating intrinsic E-mode polarization at late stages of our universe (a.k.a the cosmic birefringence effect); hence, a variety of late-time phenomena relevant to new particles beyond SM, such as axions, have been examined \cite{Carroll:1998:DE,Panda:2010,Fujita:2020aqt,Fujita:2020ecn,Choi:2021aze,Obata:2021,Gasparotto:2022uqo,Galaverni:2023,Gasparotto:2023psh,Murai:2022:EDE,Eskilt:2023:EDE,Finelli:2009,Liu:2016dcg,Fedderke:2019:biref,Takahashi:2020tqv,Kitajima:2022jzz,Jain:2022jrp,Gonzalez:2022mcx}. 
If the parity symmetry is broken by primordial gravitational waves, it would be also imprinted in the $TB$ and $EB$ correlations.
Although such a primordial tensor signal has not been detected with $>2\sigma$ level so far in these cross-correlations~\cite{Saito:2007kt,Gerbino:2016mqb,Fujita:2022qlk,Campeti:2022acx} as well as 3-point correlations (bispectra) \cite{Shiraishi:2014ila,Planck:2015zfm,Planck:2019kim}, many theoretical works have been done~\cite{sorbo:2011,Barnaby:2011vw,cook/sorbo:2012,Barnaby:2011qe,barnaby/etal:2012,cook/sorbo:2013,Mukohyama:2014gba,Bartolo:2014hwa,Bartolo:2015dga,Namba:2015gja,Domcke:2016bkh,Garcia-Bellido_2016,Obata:2016oym,Domcke:2017fix,Garcia-Bellido:2017aan,Ozsoy:2020ccy,Caravano:2021pgc,Caravano:2021bfn,Caravano:2022epk,Niu:2023bsr,Unal:2023srk,maleknejad/sheikh-jabbari:2011,Dimastrogiovanni:2012ew,Shiraishi:2013kxa,Adshead:2013nka,Adshead:2013qp,Dimastrogiovanni_fasiello_fujita_2016,Shiraishi:2016yun,adshead/sfakianakis:2017,Fujita:2017jwq,adshead/martinec/sfakianakis:2016,Agrawal_etal,Thorne,agrawal/fujita/komatsu:2018b,Papageorgiou:2018rfx,Fujita:2018vmv,Fujita:2018ndp,Papageorgiou:2019ecb,Campeti:2020xwn,Ishiwata:2021yne}.
This is because the production of the parity-violating gravitational waves is well motivated by axions.
Axion-gauge dynamics could trigger the particle production of one helicity mode of gauge field and source the cosmological perturbations  (also, see \cite{Maleknejad_etal,Komatsu:2022nvu} for recent reviews).

Differently from the tensor sector, the parity-odd information in the primordial scalar sector is inaccessible in the 2-point and 3-point statistics due to the statistical isotropy of our universe. Hence, a 4-point correlation (trispectrum) becomes the lowest-order diagonostic tool for the scalar-sector parity violation \cite{Shiraishi:2016mok}. Motivated by this fact, Refs.~\cite{Philcox:2021hbm,Cahn:2021ltp,Hou:2022wfj,Philcox:2022hkh,Cabass:2022oap,Creque-Sarbinowski:2023wmb} have measured the trispectrum of galaxy number density fields with the BOSS galaxy-clustering data and found $> 3 \sigma$ parity-odd signals. Measuring the CMB temperature and E-mode polarization trispecta using the {\it Planck} data has indicated weaker significance level \cite{Philcox:2023ffy,Philcox:2023ypl}, while it is still worth discussing the relation to new physics, and several generation mechanisms have already been proposed \cite{Shiraishi:2016mok,Liu:2019fag,Cabass:2022rhr,Cabass:2022oap,Niu:2022fki,Creque-Sarbinowski:2023wmb,Jazayeri:2023kji,Stefanyszyn:2023qov}.

In anticipation of a physical interpretation of the measured large parity-odd signal in the scalar trispectrum, this paper focuses on the axion-gauge dynamics during inflation, and, for the first time, examines the trispectrum generation in a model of spectator axion-$U(1)$ gauge field, called the rolling axion model~\cite{Namba:2015gja}. In this model, an axion is not an inflaton but a spectator field and the axion velocity dynamically evolves in time. At one-loop level, the sourced gauge field provides the scalar-mode correlation with a unique scale dependence, and generates a sizable amount of non-Gaussianity at around equilateral wavenumber configurations. Via massive analytical and numerical analyses, we confirm that the induced scalar trispectrum has not only parity-even but also parity-odd signals, which corresponds to real and imaginary numbers, respectively. We also find that the ratio of parity-odd to parity-even mode is ${\cal O}(10\%)$ at the exact equilateral limit ($k_1 = k_2 = k_3 = k_4$), and more interestingly, can reach/exceed unity for quasi-equilateral configurations ($k_1 = k_2 = k_3 \sim k_4$) depending on orientations of wavevectors. 

Previous studies~\cite{Shiraishi:2016mok,Niu:2022fki,Stefanyszyn:2023qov} have also computed the scalar trispectrum in other axion models where an inflaton is identified with an axion, and the axion velocity is constant in time.%
\footnote{For a tensor trispectrum analysis, see Ref.~\cite{Fujita:2021flu}.}
The model of Ref.~\cite{Shiraishi:2016mok} (originally \cite{Bartolo:2015dga}), predicting a sizable signal rather for collapsed configurations (e.g., $|\bm{k}_1 + \bm{k}_2| \ll k_1, k_2, k_3, k_4$) due to a $f(\phi)(F^2 + F\tilde{F})$ coupling, has already been tested with the {\it Planck} and BOSS data, and no significant evidence (with a maximal deviation of $2.0\sigma$) has been found \cite{Philcox:2022hkh,Philcox:2023ffy,Philcox:2023ypl}. 
The model of Ref.~\cite{Stefanyszyn:2023qov} includes a nonvanishing mass term of the gauge field, inducing different trispectrum shapes.
Regarding the model of Ref.~\cite{Niu:2022fki}, like our case, the trispectrum peaks for the equilateral configurations. However, due to the difference of the generation mechanism, the parity-odd signal in our model can become larger than that in Ref.~\cite{Niu:2022fki}; thus, our model may be more useful on a physical interpretation of the large parity-odd trispectrum signal observed in the BOSS data.

This paper is organized as follows.
In section~\ref{sec:model}, we 
briefly describe a model of rolling axion-$U(1)$ model.
In section~\ref{sec:production}, we present a perturbation dynamics of sourced scalar mode by the enhanced gauge field.
In section~\ref{sec:formalism}, we analytically formulate the scalar-mode trispectrum.
In section~\ref{sec:result}, we numerically evaluate the parity-odd and parity-even component in the trispectrum.
We will summarize our result in section~\ref{sec:conclusion}. 
Throughout this paper, we set a natural unit $\hbar = c = 1$.

\section{Model}
\label{sec:model}

We consider the following axion-$U(1)$ spectator model (known as a rolling axion model \cite{Namba:2015gja}) where a spectator axionic field $\sigma$ couples to $U(1)$ gauge field $A_\mu$ during inflation:
\begin{equation}
S = \int d^4x\sqrt{-g}\left[ \dfrac{1}{2}\Mpl^2 R - \dfrac{1}{2}(\partial\varphi)^2 - U(\varphi) - \dfrac{1}{2}(\partial\sigma)^2 - V(\sigma) - \dfrac{1}{4}F_{\mu\nu}F^{\mu\nu} + \dfrac{\lambda}{4}\dfrac{\sigma}{f}F_{\mu\nu}\tilde{F}^{\mu\nu} \right] \ ,
\end{equation}
where $R$ is the Ricci scalar, $\varphi$ is an inflaton, $U(\varphi)$ is its potential, $V(\sigma)$ is a potential of the axion, $F_{\mu\nu} = \partial_\mu A_\nu - \partial_\nu A_\mu$ is the field strength of $U(1)$ gauge field, $\tilde{F}^{\mu\nu} = \epsilon^{\mu\nu\rho\sigma}F_{\rho\sigma}/(2\sqrt{-g})$ is its Hodge dual, $\epsilon^{\mu\nu\rho\lambda}$ is the Levi-Civit\`{a} symbol with $\epsilon^{0123}=1$, $M_{\rm Pl} \equiv 1 / \sqrt{8 \pi G}$ is the reduced Planck mass, and $\lambda ,f$ are constant parameters. 
We consider the flat FLRW Universe, ${\rm d}s^2=-{\rm d}t^2+a^2(t){\rm d}{\bm x}^2$, with $a(t)$ being the scale factor. In what follows,
we assume that the unspecified inflaton field dominates the inflationary universe (i.e. $V(\sigma)\ll U(\varphi)$) and the Hubble parameter $H(t)\equiv \dot{a}/a$ is well approximated by a constant (i.e. $
\dot{\varphi}^2\ll U(\varphi)$), where dot denotes the time derivative, $\dot{X} \equiv\partial_t X$.
Our analysis in this and the next section mostly follows that of Ref.~\cite{Namba:2015gja}.

Regarding the spectator potential, we adopt the cosine potential form:
\begin{equation}
V(\sigma) = \Lambda^4\left[ 1 - \cos\left(\dfrac{\sigma}{f}\right) \right] \ .
\end{equation}
The equation of motion (EoM) for the background spectator field $\sigma(t)$ is given by
\begin{equation}
\ddot{\sigma} + 3H\dot{\sigma} + \partial_\sigma V = 0 \ ,
\end{equation}
where we neglect the backreaction term from the gauge field, $\lambda \langle E_i B_i\rangle/f$.
Further neglecting the $\ddot{\sigma}$ term and solving the slow-roll equation $3H\dot{\sigma} = -\partial_\sigma V$, 
we obtain the solution as
\begin{equation}
\sigma(t) = 2f\text{Arccot}\left(e^{\delta H (t-t_*)}\right) \ , \qquad \delta \equiv \dfrac{\Lambda^4}{3H^2f^2} \ ,
\end{equation}
where a dimensionless parameter $\delta$ characterizes the width of cosine potential
and $t_*$ is a time when the axion passes through the inflection point of the potential.
Then, the slow-roll condition
\begin{equation}
  \frac{\ddot{\sigma}}{3 H \dot{\sigma}}
  = -\frac{\delta}{3} \tanh{\left(\delta H (t-t_*)\right)} \ll 1,
\end{equation}
is valid when $\delta \ll 3$ is satisfied.

For later convenience, we define the dimensionless parameter
\begin{equation}
\xi(t) \equiv -\lambda\dfrac{\dot{\sigma}(t)}{2fH} > 0 \quad (\dot{\sigma} < 0)
\end{equation}
which describes the speed of spectator field
and we assume it is positive in this paper.
The slow-roll solution for $\xi$ is given by
\begin{equation}
\xi(t) = \dfrac{\xi_*}{\cosh(\delta H(t-t_*))} \ , \qquad \xi_* \equiv \dfrac{\lambda\delta}{2} \ ,
\end{equation}
where $\xi_* = \xi(t_*)$ is the maximum value of $\xi$.

In analyzing the gauge field perturbation, we switch the time variable from the cosmic time $t$ to the conformal time $\tau$.
We fix the gauge by choosing the Coulomb gauge $\partial_iA_i=0$ and the temporal gauge $A_0=0$. 
To describe the production of quanta of the gauge field, we promote the classical field $A_i$ to an operator
\begin{align}
\hat{A}_i(\tau, \bm{x}) 
&= \sum_{\lambda=\pm}\int\dfrac{{\rm d}^3 k}{(2\pi)^3}\hat{A}^\lambda_{\bm{k}}(\tau)e^\lambda_i(\hat{\bm{k}})e^{i\bm{k}\cdot\bm{x}} \ , \\
\hat{A}^\lambda_{\bm{k}}(\tau) &= \hat{a}^\lambda_{\bm{k}}A^\lambda_k(\tau) + \hat{a}^{\lambda\dagger}_{-\bm{k}}A^{\lambda*}_k(\tau)
,\quad (\lambda = \pm) \ ,
\end{align}
where 
$\hat{a}^{\lambda}_{\bm{k}},\hat{a}^{\lambda\dagger}_{\bm{k}}$ are the
creation and annihilation operators obeying the commutation relation, $[\hat{a}^\lambda_{\bm{k}}, \ \hat{a}^{\lambda'\dagger}_{-\bm{k}'}] = \delta^{\lambda\lambda'}(2\pi)^3\delta^3(\bm{k}+\bm{k}')$.
We employ the circular polarization vectors
\begin{equation}
\bm{e}^{\pm}(\hat{\bm{k}}) = \dfrac{1}{\sqrt{2}}(\cos\theta_{\hat{\bm{k}}}\cos\phi_{\hat{\bm{k}}}\mp i\sin\phi_{\hat{\bm{k}}}, \ \cos\theta_{\hat{\bm{k}}}\sin\phi_{\hat{\bm{k}}}\pm i\cos\phi_{\hat{\bm{k}}}, -\sin\theta_{\hat{\bm{k}}}) \ ,
\end{equation}
in terms of $\hat{\bm{k}} = (\sin\theta_{\hat{\bm{k}}}\cos\phi_{\hat{\bm{k}}}, \sin\theta_{\hat{\bm{k}}}\sin\phi_{\hat{\bm{k}}}, \cos\theta_{\hat{\bm{k}}})$, and it satisfies the orthonormal condition $e^{\lambda}_i(\hat{\bm{k}})e^{\lambda'*}_i(\hat{\bm{k}})=\delta^{\lambda\lambda'}$, the transverse property $k_ie^\pm_i(\hat{\bm{k}}) = 0$ and the eigenvectors of the curl $i\epsilon_{ijk}k_je^\pm_k(\hat{\bm{k}})=\pm ke^\pm_i(\hat{\bm{k}})$, where $k=|\bm{k}|$, $\hat{\bm{k}}=\bm{k}/k$ and 
$\epsilon_{ijk} \equiv \epsilon_{0ijk}$.

The EoM for the mode function of the gauge field reads
\begin{equation}
\left[ \partial_\tau^2 + k^2 \pm \dfrac{2k\xi(\tau)}{\tau}\right]A^\pm_k(\tau) = 0 \ , \qquad \xi(\tau) = \dfrac{2\xi_*}{\left(\tfrac{\tau_*}{\tau}\right)^\delta + \left(\tfrac{\tau}{\tau_*}\right)^\delta} \ , \label{eq: eomA}
\end{equation}
where we use $a=-1/(H\tau)$ and $a_*=-1/(H\tau_*)$.
While the minus mode $A^-_k$ ~is not amplified, we can see that the plus mode $A^+_k$ experiences a tachyonic instability for $-k\tau < 2\xi(\tau)$.
We are interested in the parameter region $\xi(\tau) \gtrsim 1$, so that $A^+_k$ starts to grow slightly before the horizon-crossing, $-k\tau=1$. 

Although the solution of \eqref{eq: eomA} cannot be written in a closed form, we can use the WKB approximation.
Then, for $\tau > \bar{\tau}$ obeying $-k\bar{\tau} = 2\xi(\bar{\tau})$, we obtain the following approximate solution:
\begin{align}
A^+_k(\tau) &\simeq 
\dfrac{N(\xi_*, -k\tau_*, \delta)}{\sqrt{2k}}\left(\dfrac{-k\tau}{2\xi(\tau)}\right)^{1/4}
\notag\\
&\times\exp\left[ -\dfrac{4\xi_*^{1/2}}{1+\delta}\left(\dfrac{\tau}{\tau_*}\right)^{\delta/2}(-k\tau)^{1/2}{_2}F_1\left(\tfrac{1}{2}, \ \tfrac{1+\delta}{4\delta}; \ \tfrac{5\delta+1}{4\delta}; \ -\left(\tfrac{\tau}{\tau_*}\right)^{2\delta}\right) \right]\,,
\notag \\
 &\equiv \dfrac{1}{\sqrt{2k}}\mathcal{A}(\xi, -k\tau) \ , \label{eq: A}\\
\partial_\tau A^+_k(\tau) &\simeq \sqrt{\dfrac{2k\xi(\tau)}{-\tau}}A^+_k(\tau) \label{eq: A'} \ ,
\end{align}
where the normalization factor $N$ is determined by matching $A^+_k$ at late time to the full numerical solution of \eqref{eq: eomA} and ${_2}F_1$ is a hypergeometric function.
In the deep super-horizon regime $-k\tau \rightarrow 0$, the energy density of gauge fields dilutes due to the expansion of the universe.
Thus, the gauge field makes the most contribution to the physical enhancement of fluctuations at around the horizon-crossing.
A similar approximated expression was used in the previous work~\cite{Namba:2015gja}. Here we find a more accurate expression with the hypergeometric function, which is derived in Appendix \ref{app: WKB}.

\section{Scalar mode production}\label{sec:production}

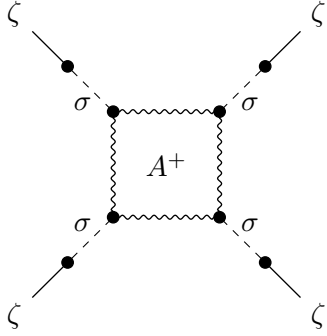
\begin{figure}[thpb]    
\begin{center}     
\begin{tikzpicture} 
\begin{feynhand}
    \vertex [particle] (i1) at (-2,2) {$\zeta$};
    \vertex [particle] (i2) at (-2,-2) {$\zeta$};
    \vertex [particle] (i3) at (2,-2) {$\zeta$};
    \vertex [particle] (i4) at (2,2) {$\zeta$};
    \vertex [particle] (i5) at (-1.1,0.8) {$\sigma$};
    \vertex [particle] (i6) at (-1.1,-0.8) {$\sigma$};
    \vertex [particle] (i7) at (1.1,-0.8) {$\sigma$};
    \vertex [particle] (i8) at (1.1,0.8) {$\sigma$};
    \vertex [particle] (i9) at (0,0) {$A^+$};
    \vertex [dot] (v1) at (-1.3,1.3) {};
    \vertex [dot] (v2) at (-1.3,-1.3) {};
    \vertex [dot] (v3) at (1.3,-1.3) {};
    \vertex [dot] (v4) at (1.3,1.3) {};
    \vertex [dot] (w3) at (-0.7,0.7) {};
    \vertex [dot] (w4) at (0.7,0.7) {};
    \vertex [dot] (w5) at (-0.7,-0.7) {};
    \vertex [dot] (w6) at (0.7,-0.7) {};
    \propag [plain] (i1) to (v1);
    \propag [plain] (i2) to (v2);
    \propag [plain] (i3) to (v3);
    \propag [plain] (i4) to (v4);
    \propag [dashed] (v1) to (w3);
    \propag [dashed] (v2) to (w5);
    \propag [dashed] (v3) to (w6);
    \propag [dashed] (v4) to (w4);
    \propag [photon] (w3) to (w4);
    \propag [photon] (w3) to (w5);
    \propag [photon] (w5) to (w6);
    \propag [photon] (w4) to (w6);
    \end{feynhand}
\end{tikzpicture}
\caption{Diagram of the four-point function that we compute. The external solid line represents the curvature perturbation $\zeta$. The internal dashed line shows the axion $\sigma$. The intermediate wiggly line represents the one helicity of gauge field $A^+$.}
\label{fig:diagram}
\end{center}
\end{figure}

We will estimate the four-point correlation function of the curvature perturbation on a flat hypersurfaces $\zeta = -H\delta\varphi/\dot{\varphi}$ induced by the gauge field perturbation, which is depicted by a one-loop diagram in Figure \ref{fig:diagram}.
To this end, we first calculate the fluctuation of the spectator axion $\delta\sigma$ in this section. 
$\delta\sigma$ is produced by the intrinsic inhomogeneity of the Chern-Simons gauge interaction $\delta_{E B} \equiv \left( E_i B_i - \langle E_i B_i \rangle \right)|_{\delta\sigma=0}$, where $E_i = -\partial_\tau A_i/a^2, \ B_i = \epsilon_{ijk}\partial_jA_k/a^2$.
 The EoM for $\delta\sigma$ is given by
\begin{align}
\ddot{\delta\sigma} + 3H\dot{\delta\sigma} - a^{-2}\nabla^2\delta\sigma \simeq -\dfrac{\lambda}{f}\delta_{E B} \label{eq: deltavarphii} \ .
\end{align}
Note that we do not include the mass term of $\delta\sigma$ and other higher order interactions for simplicity. 
The effective mass squared of $\delta\sigma$ is $V''(\sigma)=3\delta H^2 \cos{(\sigma /f)}$, it exactly vanishes at the inflection point, and it remains small in other regions under the the slow-roll condition $\delta \ll 3$.

Since $\langle E_i B_i \rangle(t)$ contributes only to zero-mode, the Fourier transform of the above EoM reads
\begin{equation}
\left[ \dfrac{\partial^2}{\partial \tau^2} + k^2 - \dfrac{2}{\tau^2} \right](a\delta\hat{\sigma}_{\bm{k}}) = \int \dfrac{{\rm d}^3 p}{(2\pi)^3}\hat{\mathcal{S}}(\tau, \bm{p}, \ \bm{k}-\bm{p}) \ ,
\end{equation}
where
\begin{equation}
\hat{\mathcal{S}}(\tau, \ \bm{p}, \ \bm{k}-\bm{p}) \equiv -\lambda\dfrac{a^3}{2f}e^+_i(\hat{\bm{p}})e^+_i(\reallywidehat{\bm{k} - \bm{p}})\left( \hat{E}^+_{\bm{p}} \hat{B}^+_{\bm{k} - \bm{p}} + \hat{E}^+_{\bm{k} - \bm{p}}\hat{B}^+_{\bm{p}} \right),
\end{equation}
where we define $\hat{E}^{+}_{\bm{p}}=-\partial_\tau \hat{A}^{+}_{\bm{p}}/a^2,$ 
and $\hat{B}^{+}_{\bm{p}}=|\bm{p}|\hat{A}^{+}_{\bm{p}}/a^2$,
and we have symmetrized the operators of the source term with respect to the momenta $\bm{p}$ and $\bm{k}-\bm{p}$ to make the correlation functions independent from the order of operators.
$\delta\sigma_k$ has two solutions $\delta\sigma_k = \delta\sigma^{\rm (V)}_k + \delta\sigma^{\rm (S)}_k$. 
The homogeneous solution ~$\delta\sigma^{\rm (V)}_k$ ~is a normal vacuum fluctuation and its expression for the Bunch-Davies initial condition is given by the first kind of Hankel function $H_\nu^{(1)} = J_\nu + iY_\nu$:
\begin{equation}
a\delta\sigma_{k}^{\rm (V)} = -\dfrac{1}{\sqrt{2k}}\sqrt{\dfrac{-\pi k\tau}{2}}H^{(1)}_{3/2}(-k\tau) \equiv u(-k\tau) \ .
\label{vac dsigma}
\end{equation}
In the super-horizon limit, this solution yields the well-known amplitude $\langle (\delta\sigma_k^{\rm (V)})^2 \rangle^{1/2} = H/(2\pi)$.
In contrast, the inhomogeneous solution $\delta \sigma_k^{\rm (S)}$ sourced by the gauge field can be found as
\begin{align}
\delta\hat{\sigma}_{k}^{\rm (S)} &\simeq \dfrac{1}{a}\int_{-\infty}^{\infty} d\tau'G_k(\tau, \tau')\int\dfrac{{\rm d}^3 p}{(2\pi)^3} \hat{\mathcal{S}}(\tau', \ \bm{p}, \ \bm{k}-\bm{p}) \label{eq: s} \ ,
\end{align}
where 
\begin{align}
G_k(\tau,\tau') &= i\Theta(\tau-\tau')\left[u(-k\tau)u^*(-k\tau') - u^*(-k\tau)u(-k\tau')\right], \notag \\
&= \Theta(\tau-\tau')\dfrac{\pi}{2}\sqrt{\tau\tau'}\left[ J_{3/2}(-k\tau)Y_{3/2}(-k\tau') - Y_{3/2}(-k\tau)J_{3/2}(-k\tau') \right],
\label{Greens function}
\end{align}
is a retarded Green function and in the super-horizon limit of $\tau \rightarrow 0$ it reads
\begin{equation}
G_k(\tau \rightarrow 0, \tau') = \Theta(\tau-\tau')\sqrt{\dfrac{\pi}{2}}\dfrac{\sqrt{-k\tau'}}{-k^{2}\tau}J_{3/2}(-k\tau') \ .
\end{equation}
%
%

With the aid of slow-roll parameters of the background inflaton and the axion,
\begin{equation}
\epsilon_\varphi \equiv \dfrac{\dot{\varphi}^2}{2\Mpl^2H^2} \ ,\qquad \epsilon_\sigma \equiv \dfrac{\dot{\sigma}^2}{2\Mpl^2H^2} \ ,
\end{equation}
the EoM for the fluctuation of the inflaton can be written as
\begin{equation}
\left[ \dfrac{\partial^2}{\partial \tau^2} + k^2 - \dfrac{2}{\tau^2} \right]\left(a\delta\hat{\varphi}_{\bm{k}}^{(S)}\right) \simeq \dfrac{6}{\tau^2}\sqrt{\epsilon_\varphi\epsilon_\sigma}\left(a\delta\hat{\sigma}_{\bm{k}}^{(S)}\right) \ .
\end{equation}
The source term comes from the gravitational coupling between the inflaton and the axion \cite{Namba:2015gja}.%
\footnote{Another effect, such as $A + A \to \delta \phi$, that contributes directly via gravitational coupling from the gauge field is also possible.
As shown in Ref.~\cite{barnaby/etal:2012}, the gauge field $A$ sources $\delta \phi$ only around the horizon crossing.
In contrast, the contribution sourced via $\delta \sigma$ sources $\delta \phi$ even in the super-horizon.
Therefore, the gauge field's direct contribution via gravitational coupling is smaller than the contribution sourced via $\delta \sigma$.
(See also Refs.~\cite{Namba:2015gja, Campeti:2022acx}.)}
Here we ignore the mass of the inflaton and its retarded Green function is identical to that of the axion, Eq.~\eqref{Greens function}.
Then, we obtain the inhomogeneous solution of the above equation as
\begin{equation}
\delta\hat{\varphi}_{\bm{k}}^{\rm (S)} = \dfrac{6\sqrt{\epsilon_\varphi}}{a}\int_{-\infty}^{\infty} d\tau'\dfrac{\sqrt{\epsilon_\sigma(\tau')}}{\tau{'}^2}G_k(\tau, \tau')\int_{-\infty}^{\infty} d\tau''G_k(\tau', \tau'')\int\dfrac{{\rm d}^3 p}{(2\pi)^3} \hat{\mathcal{S}}(\tau'', \ \bm{p}, \ \bm{k}-\bm{p}) \ ,
\label{sourced dphi}
\end{equation}
where we have assumed $\epsilon_\varphi = \text{const}$. for simplicity.
The curvature perturbation can be decomposed into two contributions
$\hat{\zeta}_{\bm{k}}=\zeta_{\bm{k}}^{(V)}+\hat{\zeta}_{\bm{k}}^{(S)}$, where the vacuum contribution and the sourced contribution is given by
\begin{align}
  \zeta_{\bm{k}}^{(V)}
  = -\frac{H}{\dot{\varphi}} \delta\varphi_{\bm{k}}^{(V)},
  \qquad
  \hat{\zeta}_{\bm{k}}^{(S)}
  = -\frac{H}{\dot{\varphi}} \delta\hat{\varphi}_{\bm{k}}^{(S)},
  \label{eq:zetaVS}
\end{align}
respectively.

In passing, we quickly discuss the power spectrum of curvature perturbation $\mathcal{P}_\zeta$, which is defined by
\begin{equation}
\langle \hat{\zeta}_{\bm{k}}\hat{\zeta}_{\bm{k}'} \rangle = (2\pi)^3 \delta(\bm{k} + \bm{k}')\dfrac{2\pi^2}{k^3}\mathcal{P}_\zeta\,.
\end{equation}
The contributions comes from the vacuum fluctuation $\delta \varphi_k^{(V)}$, whose mode function is identical to Eq.~\eqref{vac dsigma} in the slow-roll regime, and the sourced one is obtained in Eq.~\eqref{sourced dphi}. Combining them and using Eq.~\eqref{eq:zetaVS}, one finds
\begin{equation}
\mathcal{P}_\zeta = \mathcal{P}_{\zeta,v}\left[ 1 + \mathcal{P}_{\zeta,v}f_2(\xi)e^{4\pi\xi} \right] \ , \label{eq: power}
\end{equation}
where $\mathcal{P}_{\zeta,v} \equiv H^2/(8\pi^2\Mpl^2 \epsilon_\varphi)$ is the dimensionless power spectrum of the vacuum curvature perturbation.
The function $f_2(\xi)$ was found numerically.
For large and constant $\xi$, it is given by \cite{Barnaby:2011vw}
\begin{equation}
f_2(\xi) \simeq \dfrac{7.5\times10^{-5}}{\xi^6} \ .
\end{equation}
When $\xi$ is dynamical, a more elaborated fitting function is known~\cite{Namba:2015gja}.
One can also compute the bispectrum of the curvature perturbation induced by the sourced inflaton fluctuation $\delta\hat{\varphi}_{k}^{\rm (S)}$ in the same way~\cite{Namba:2015gja}. However, we will skip it and calculate the trispectrum in the next section.

\section{Formalism of the trispectrum}\label{sec:formalism}

In this section, we derive some expressions that will be used for the evaluation of the trispectrum in the next section. The trispectrum of the sourced curvature perturbation $T_\zeta$ is defined as
\begin{equation}
\langle \hat{\zeta}_{\bm{k}_1}^{(S)}\hat{\zeta}_{\bm{k}_2}^{(S)}\hat{\zeta}_{\bm{k}_3}^{(S)}\hat{\zeta}_{\bm{k}_4}^{(S)} \rangle \equiv (2\pi)^{3}\delta(\bm{k}_1+\bm{k}_2+\bm{k}_3+\bm{k}_4)T_\zeta(\bm{k}_1, \ \bm{k}_2, \ \bm{k}_3, \ \bm{k}_4) \ .
\end{equation}
In particular, we will calculate the real and imaginary parts of the above trispectrum separately for the following reason~\cite{Shiraishi:2016mok}.
The parity transformation in real space is given by flipping the sign of the spatial coordinates as $\bm{x}\to -\bm{x}$.
Similarly, the parity transformation in Fourier space corresponds to flipping the sign of the wave number vector as $\bm{k}\to -\bm{k}$.
Then, we consider the parity transformation of the trispectrum of curvature perturbation, we obtain
\begin{align}
  \langle
  \hat{\zeta}_{\bm{k}_1}^{(S)}
  \hat{\zeta}_{\bm{k}_2}^{(S)}
  \hat{\zeta}_{\bm{k}_3}^{(S)}
  \hat{\zeta}_{\bm{k}_4}^{(S)}
  \rangle
  \ \to \
  \langle
  \hat{\zeta}_{-\bm{k}_1}^{(S)}
  \hat{\zeta}_{-\bm{k}_2}^{(S)}
  \hat{\zeta}_{-\bm{k}_3}^{(S)}
  \hat{\zeta}_{-\bm{k}_4}^{(S)}
  \rangle
  = \langle
  \hat{\zeta}_{\bm{k}_1}^{(S)}
  \hat{\zeta}_{\bm{k}_2}^{(S)}
  \hat{\zeta}_{\bm{k}_3}^{(S)}
  \hat{\zeta}_{\bm{k}_4}^{(S)}
  \rangle^* \ ,
\end{align}
where we use the reality condition of curvature perturbation
$\zeta_{-\bm{k}}^{(S)}=\zeta_{\bm{k}}^{(S)*}$ in the last equality.
We can see that the real part of the trispectrum ${\rm Re}[T_{\zeta}]$ is invariant under parity transformation, so this is parity even.
In contrast, the imaginary part ${\rm Im}[T_{\zeta}]$ changes the sign in the parity transformation.
Therefore, the imaginary part of trispectrum is sensitive to parity violation.

We begin with the four-point correlation function of the sourced inflaton fluctuation $\delta\hat{\varphi}_{\bm{k}}^{(S)}$. Using Eq.~\eqref{sourced dphi}, we find
\begin{align}
&\langle \delta\hat{\varphi}_{\bm{k}_1}^{(S)}\delta\hat{\varphi}_{\bm{k}_2}^{(S)}\delta\hat{\varphi}_{\bm{k}_3}^{(S)}\delta\hat{\varphi}_{\bm{k}_4}^{(S)} \rangle \notag \\
 &= \dfrac{6^4\epsilon_\varphi^2}{a^4}\prod_{i=1,2,3,4}\int_{-\infty}^\infty d\tau'_i \dfrac{\sqrt{\epsilon_\sigma(\tau'_i)}}{\tau{'}_i^2}G_{k_i}(\tau, \tau'_i)\int_{-\infty}^\infty d\tau''_iG_{k_i}(\tau'_i, \tau''_i)\int\dfrac{d\bm{p}_i}{(2\pi)^3} \notag \\
 &\times \langle \hat{\mathcal{S}}(\tau''_1, \ \bm{p}_1, \ \bm{k}_1 - \bm{p}_1)\hat{\mathcal{S}}(\tau''_2, \ \bm{p}_2, \ \bm{k}_2 - \bm{p}_2)\hat{\mathcal{S}}(\tau''_3, \ \bm{p}_3, \ \bm{k}_3 - \bm{p}_3)\hat{\mathcal{S}}(\tau''_4, \ \bm{p}_4, \ \bm{k}_4 - \bm{p}_4) \rangle \ .
\end{align}
Dropping the disconnected contributions and keeping the connected ones corresponding to the Feynman diagram shown in Fig.~\ref{fig:diagram}, we obtain
\begin{align}
&\langle \hat{\mathcal{S}}(\tau''_1, \ \bm{p}_1, \ \bm{k}_1 - \bm{p}_1)\hat{\mathcal{S}}(\tau''_2, \ \bm{p}_2, \ \bm{k}_2 - \bm{p}_2)\hat{\mathcal{S}}(\tau''_3, \ \bm{p}_3, \ \bm{k}_3 - \bm{p}_3)\hat{\mathcal{S}}(\tau''_4, \ \bm{p}_4, \ \bm{k}_4 - \bm{p}_4) \rangle \notag \\
&= 2^4(2\pi)^{12}\left(\delta(\bm{p}_1+\bm{p}_2)\delta(\bm{p}_3+\bm{p}_4)\delta(\bm{k}_1-\bm{p}_1+\bm{k}_4-\bm{p}_4)\delta(\bm{k}_2-\bm{p}_2+\bm{k}_3-\bm{p}_3) \right. \notag \\
&\left. +\delta(\bm{p}_1+\bm{p}_2)\delta(\bm{p}_3+\bm{p}_4)\delta(\bm{k}_1-\bm{p}_1+\bm{k}_3-\bm{p}_3)\delta(\bm{k}_2-\bm{p}_2+\bm{k}_4-\bm{p}_4) \right. \notag \\
&\left. +\delta(\bm{p}_1+\bm{p}_3)\delta(\bm{p}_2+\bm{p}_4)\delta(\bm{k}_1-\bm{p}_1+\bm{k}_4-\bm{p}_4)\delta(\bm{k}_2-\bm{p}_2+\bm{k}_3-\bm{p}_3) \right) \notag \\
&\times \mathcal{S}(\tau''_1, \ \bm{p}_1, \ \bm{k}_1 - \bm{p}_1)\mathcal{S}(\tau''_2, \ \bm{p}_2, \ \bm{k}_2 - \bm{p}_2)\mathcal{S}(\tau''_3, \ \bm{p}_3, \ \bm{k}_3 - \bm{p}_3)\mathcal{S}(\tau''_4, \ \bm{p}_4, \ \bm{k}_4 - \bm{p}_4) \ ,
\end{align}
where the coefficient $2^4$ comes from the symmetric computation of the quadratic operator $\hat{\mathcal{S}}$.
To make the computation simple, in the whole integral range we adopt the approximation function \eqref{eq: A} and \eqref{eq: A'} which receives its support around the horizon crossing.
This approximation is valid because the UV contribution from the deep sub-horizon regime should be removed and the energy density of electromagnetic field dilutes at the deep super-horizon regime as discussed in Appendix.~\ref{app: WKB}.
Plugging Eqs.~\eqref{eq: A} and \eqref{eq: A'}, we obtain
\begin{align}
&\mathcal{S}(\tau_1, \ \bm{p}_1, \ \bm{k}_1 - \bm{p}_1)\mathcal{S}(\tau_2, \ \bm{p}_2, \ \bm{k}_2 - \bm{p}_2)\mathcal{S}(\tau_3, \ \bm{p}_3, \ \bm{k}_3 - \bm{p}_3)\mathcal{S}(\tau_4, \ \bm{p}_4, \ \bm{k}_4 - \bm{p}_4) \notag \\
&\simeq \dfrac{\lambda^4}{2^6f^4}\prod_{i=1,2,3,4}\dfrac{e^+_m(\hat{\bm{p}}_i)e^+_m(\reallywidehat{\bm{k}_i-\bm{p}_i})}{a(\tau_i)}\sqrt{\dfrac{\xi(\tau_i)}{-\tau_i}}\mathcal{A}(\xi,-p_i\tau_i)\mathcal{A}(\xi,-|\bm{k}_i-\bm{p}_i|\tau_i)\left(\sqrt{|\bm{k}_i-\bm{p}_i|} + \sqrt{p_i}\right) .
\end{align}
We introduce dimensionless time variables $x \equiv -k_1\tau, \ x_i \equiv -k_1\tau_i$ and dimensionless momenta $\bm{k}_i^* \equiv \bm{k}_i/k_1, \ \bm{p}_i^* \equiv \bm{p}_i/k_1$ normalized by $k_1$.
Note that these upper asterisks should be distinguished from the lower asterisks (e.g. 
$\xi_*\equiv \xi(t_*)$) denoting the time when the axion passes through the inflection point. Using them, we obtain
\begin{align}
&\langle \delta\hat{\varphi}_{\bm{k}_1}^{(S)}\delta\hat{\varphi}_{\bm{k}_2}^{(S)}\delta\hat{\varphi}_{\bm{k}_3}^{(S)}\delta\hat{\varphi}_{\bm{k}_4}^{(S)} \rangle
\notag \\
 &\simeq \dfrac{6^4\epsilon_\varphi^2\lambda^4H^8x^4}{4k_1^{12}f^4}\prod_{i=1,2,3,4}\Bigg(\int_{\infty}^{-\infty}dx'_i\dfrac{\sqrt{\epsilon_\sigma(\tau_i')}}{x_i{'}^2}k_1G_{k_i}(x,x_i')\int_{\infty}^{-\infty}dx''_i\sqrt{\dfrac{\xi(\tau_i'')}{x''_i}}k_1x_i''G_{k_i}(x_i',x_i'') \notag \\
&\times\int\dfrac{d\bm{p}^*_i}{(2\pi)^3}e^+_m(\hat{\bm{p}_i})e^+_m(\reallywidehat{\bm{k}_i-\bm{p}_i})\left(\sqrt{p_i^*}+ \sqrt{|\bm{k}_i - \bm{p}_i|^*}\right)\mathcal{A}(\xi, p^*_ix''_i)\mathcal{A}(\xi, |\bm{k}_i-\bm{p}_i|^*x''_i)\Bigg) \notag \\
 &\times (2\pi)^{12}\left[\delta(\bm{p}^*_1+\bm{p}^*_2)\delta(\bm{p}^*_3+\bm{p}^*_4)\delta(\bm{k}^*_1-\bm{p}_1^*+\bm{k}_4^*-\bm{p}_4^*)\delta(\bm{k}_2^*-\bm{p}_2^*+\bm{k}_3^*-\bm{p}_3^*) \right. \notag \\
 &\left.
 + (\text{permutations of} \ \bm{k}_3^*-\bm{p}_3^*, \ \bm{k}_4^*-\bm{p}_4^*) + (\text{permutations of} \ \bm{p}_2^*, \ \bm{p}_3^*)\right] \ .
 \label{dphi four}
\end{align}
Since $\delta\hat{\varphi}_{\bm{k}}^{(S)}$ and $\hat{\zeta}_{\bm{k}}^{(S)}$ are related by Eq.~\eqref{eq:zetaVS}, the trispectrum $T_\zeta$ can be calculated from Eq.~\eqref{dphi four}.
By computing the integration with delta functions and using $H/\dot{\varphi} = -1/(\Mpl\sqrt{2\epsilon_\varphi})$ and $\sqrt{\epsilon_\sigma(\tau_i')} = \sqrt{2}f\xi(\tau_i')/(\lambda\Mpl)$, we find
the asymptotic form of super-horizon limit $x\rightarrow0$ as
\begin{align}
&T_\zeta(\bm{k}_1, \ \bm{k}_2, \ \bm{k}_3, \ \bm{k}_4) \notag \\
 &\simeq  \dfrac{3^4}{2^4}\pi^8r_v^4\dfrac{\mathcal{P}_{\zeta,v}^4}{k_1^9}\int\dfrac{d\bm{p}^*_1}{(2\pi)^3}\prod_{i=1,2,3,4}e^+_m(\hat{\bm{p}_i})e^+_m(\reallywidehat{\bm{k}_i-\bm{p}_i})\left(\sqrt{p_i{^*}}+ \sqrt{|\bm{k}_i - \bm{p}_i|^*}\right)\left(p^*_i|\bm{k}_i-\bm{p}_i|^*\right)^{1/4} \notag \\
&\times \mathcal{T}\left(\xi_*, x_*, \delta, \sqrt{p_i{^*}} + \sqrt{|\bm{k}_i-\bm{p}_i|^*}\right)N(\xi_*,p^*_ix_*, \delta)N(\xi_*,|\bm{k}_i-\bm{p}_i|^*x_*, \delta) \notag \\
&+ (\text{permutations of} \ \bm{k}_3^*-\bm{p}_3^*, \ \bm{k}_4^*-\bm{p}_4^*) + (\text{permutations of} \ \bm{p}_2^*, \ \bm{p}_3^*) \ . \label{eq: tri}
\end{align}
where $r_v = 16\epsilon_\varphi$ is a vacuum tensor-to-scalar ratio and in the first line the momenta are fixed as $\bm{p}_2 = -\bm{p}_1, \ \bm{p}_3 = \bm{k}_2 + \bm{k}_3 + \bm{p}_1, \ \bm{p}_4 = \bm{k}_1 + \bm{k}_4 - \bm{p}_1$.
Here, $\mathcal{T}$ denotes the function with time integration
\begin{align}
&\mathcal{T}\left(\xi_*, x_*, \delta, \sqrt{p_i{^*}} + \sqrt{|\bm{k}_i-\bm{p}_i|^*}\right) \notag \\
&\equiv  \int_{\infty}^{0}\dfrac{dx'_i}{x_i'}\dfrac{2\xi_*}{\left(\tfrac{x_*}{x_i'}\right)^\delta + \left(\tfrac{x_i'}{x_*}\right)^\delta}\sqrt{\dfrac{\pi}{2}}\dfrac{J_{3/2}(k_i^*x_i')}{(k_i^{*}x_i{'})^{3/2}} \notag \\
&\times\int_{\infty}^{x_i'}dx''_i (x_i'x_i'')^{3/2}\dfrac{\pi}{2}\left[ J_{3/2}(k_i^*x_i')Y_{3/2}(k_i^*x_i'') - Y_{3/2}(k_i^*x_i')J_{3/2}(k_i^*x_i'') \right] \notag \\
&
\times \exp\left[ -\dfrac{4\sqrt{\xi_*x''_i}}{1+\delta}\left(\dfrac{x''_i}{x_*}\right)^{\delta/2}\left(\sqrt{p_i{^*}} + \sqrt{|\bm{k}_i-\bm{p}_i|^*}\right){_2}F_1\left(\tfrac{1}{2}, \ \tfrac{1+\delta}{4\delta}; \ \tfrac{5\delta+1}{4\delta}; \ -\left(\tfrac{x_i''}{x_*}\right)^{2\delta}\right) \right] \ .
\end{align}
In the above integration, we can find that the time integration can cover most of the interval when $p_i^* \sim |\bm{k}_i - \bm{p}_i|^* = \mathcal{O}(1)$ at which the two $\mathcal{A}$'s shown in Eq.~\eqref{dphi four} are simultaneously at their peaks.
Therefore, the trispectrum is expected to be mostly amplified at around the equilateral limit: $k_1 = k_2 = k_3 = k_4$. Hereafter, we evaluate the trispctrum signal around there.
\begin{figure}[htpb]
\begin{center}
    \includegraphics[clip, width=0.80\columnwidth]{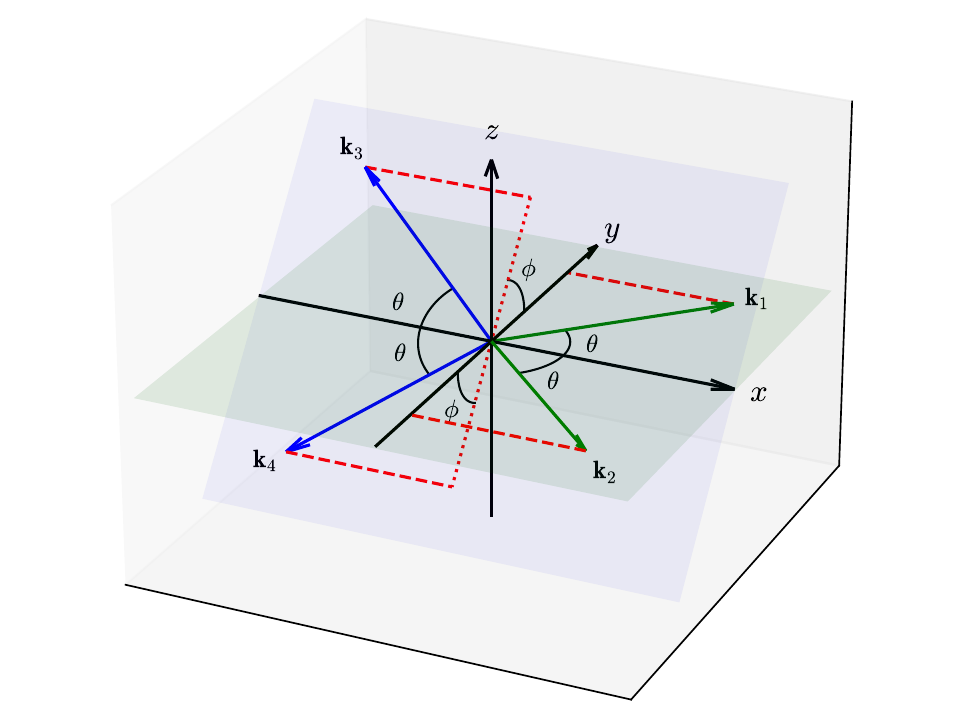}
\caption{
Our momentum configuration in the equilateral limit.
The four momenta have the same length, $k_1 = k_2 = k_3 = k_4$.
$\bm{k}_1$ and $\bm{k}_2$ are in the $x\text{-}y$ plane (green plane).  
$\bm{k}_3$ and $\bm{k}_4$ lie in another plane at an angle of $\phi$ to the $x\text{-}y$ plane around the $x$-axis (blue plane).
$\bm{k}_1+\bm{k}_2$ and $-\bm{k}_3-\bm{k}_4$ are parallel to the $x$-axis. Each of $\bm{k}_1,\bm{k}_2,-\bm{k}_3$ and $-\bm{k}_4$ makes an angle of $\theta$ with the $x$ axis.
}
\label{fig: equilateral}
\end{center}
\end{figure}

\section{Results}\label{sec:result}

In this section, we provide the numerical results of the trispectrum of the sourced curvature perturbation and discuss its parity violation.
We consider two different configurations of the four momenta; the exact equilateral and quasi-equilateral shapes.

\subsection{Exact equilateral shape}

To investigate the trispectrum, we consider four momenta $\bm{k}_1$, $\bm{k}_2$, $\bm{k}_3$, and $\bm{k}_4$ satisfying $\bm{k}_1 + \bm{k}_2 + \bm{k}_3 + \bm{k}_4 = \bm{0}$.
Note that we cannot generally choose the coordinate system where all of the momenta lie in the $x$-$y$ plane unlike the case of power spectrum and bispectrum.
We first consider the exact equilateral limit, $k_1 = k_2 = k_3 = k_4 \equiv k$, and use the parameterization of \cite{Fujita:2021flu} as shown in Fig.~\ref{fig: equilateral}:
\begin{align}
    \bm{k}_1 
    &=
    k (\cos \theta, \, \sin \theta, 0),
    \\
    \bm{k}_2 
    &=
    k (\cos \theta, \, -\sin \theta, 0),
    \\
    \bm{k}_3 
    &=
    k (-\cos \theta , \, \sin \theta \cos \phi, \, \sin \theta \sin \phi), 
    \\
    \bm{k}_4
    &=
    k (-\cos \theta , \, -\sin \theta \cos \phi, \, -\sin \theta \sin \phi).
\end{align}
In the equilateral limit, we can always take this coordinate system by setting $\hat{\bm{x}} \parallel \bm{k}_1 + \bm{k}_2$ and $\hat{\bm{z}} \perp \bm{k}_1$.
Without loss of generality, we can set $0 < \theta < \pi/2$ and $0 \leq \phi < \pi$ by exchanging $\bm{k}_i$.
Note that this coordinate system avoids $\bm{k}_i \parallel \pm \hat{\bm{z}}$, where the phases of polarization vector and tensor are not well-defined.

In the equilateral configuration, the trispectrum $T_\zeta(\bm{k}_1, \ \bm{k}_2, \ \bm{k}_3, \ \bm{k}_4)$ in
Eq.~\eqref{eq: tri} is labeled by the two angles $(\theta, \phi)$
and thus we call it $T^{(\theta, \phi)}_{\zeta,\rm eq}(k)$.
Then we normalize it as
\begin{equation}
T^{(\theta, \phi)}_{\zeta,\rm eq}(k) = P_{\zeta,v}(k)^3f^{(\theta, \phi)}_{\zeta,4}(k/k_*, \xi_*, \delta) \ ,
\label{T and f}
\end{equation}
where
the power spectrum of the vacuum curvature perturbation, $P_{\zeta,v} = 2\pi^2\mathcal{P}_{\zeta,v}/k^3$, is used. The dimensionless trispectrum is given by
\begin{align}
&f^{(\theta, \phi)}_{\zeta,4}(k/k_*, \xi_*, \delta) \notag \\
 &=  \dfrac{3^4\pi^2r_v^4}{2^7}\mathcal{P}_{\zeta,v}\int\dfrac{d\bm{p}^*_1}{(2\pi)^3}\prod_{i=1,2,3,4}e^+_m(\hat{\bm{p}_i})e^+_m(\reallywidehat{\bm{k}_i-\bm{p}_i})\left(\sqrt{p_i{^*}}+ \sqrt{|\bm{k}_i - \bm{p}_i|^*}\right)\left(p^*_i|\bm{k}_i-\bm{p}_i|^*\right)^{1/4} \notag \\
&\times \mathcal{T}\left(\xi_*, x_*, \delta, \sqrt{p_i{^*}} + \sqrt{|\bm{k}_i-\bm{p}_i|^*}\right)N(\xi_*,p^*_ix_*, \delta)N(\xi_*,|\bm{k}_i-\bm{p}_i|^*x_*, \delta) \notag \\
&+ (\text{permutations of} \ \bm{k}_3^*-\bm{p}_3^*, \ \bm{k}_4^*-\bm{p}_4^*) + (\text{permutations of} \ \bm{p}_2^*, \ \bm{p}_3^*) \ .
\label{eq:f_zeta}
\end{align}
We numerically compute this factor. Before taking a closer look at it, however, we first take an overview of its magnitude.

To characterize the trispectrum averaged with respect to the angular variables, we introduce an averaged
$f^{(\theta, \phi)}_{\zeta,4}$ as
\begin{equation}
\bar{f}_{\zeta,4}(k/k_*, \xi_*, \delta)
\equiv P_{\zeta,v}(k)^{-3} \sqrt{\dfrac{1}{N_{\rm bin}}\sum_{i, j}
\left[T^{(\theta_i, \phi_j)}_{\zeta, \rm eq}(k)\right]^2}\,,
\end{equation}
where
we have discretized an angular space $(\theta, \phi)$ into $(\theta_i, \phi_j)$ with $(i=1, ..., i_{\rm max}, \ j=1, ..., j_{\rm max})$.
$N_{\rm bin} = i_{\rm max} \times j_{\rm max}$ is a number of bins that we numerically evaluate.
We separate $\bar{f}_{\zeta,4}$ into real and imaginary parts:
\begin{equation}
\bar{f}_{\zeta,4}(k/k_*, \xi_*, \delta) \equiv F_{\zeta,4}(k/k_*, \xi_*, \delta) + i G_{\zeta,4}(k/k_*, \xi_*, \delta) \ .
\label{F&G def}
\end{equation}

\begin{figure}[htpb]
\begin{center}
    \includegraphics[clip, width=0.49\columnwidth]{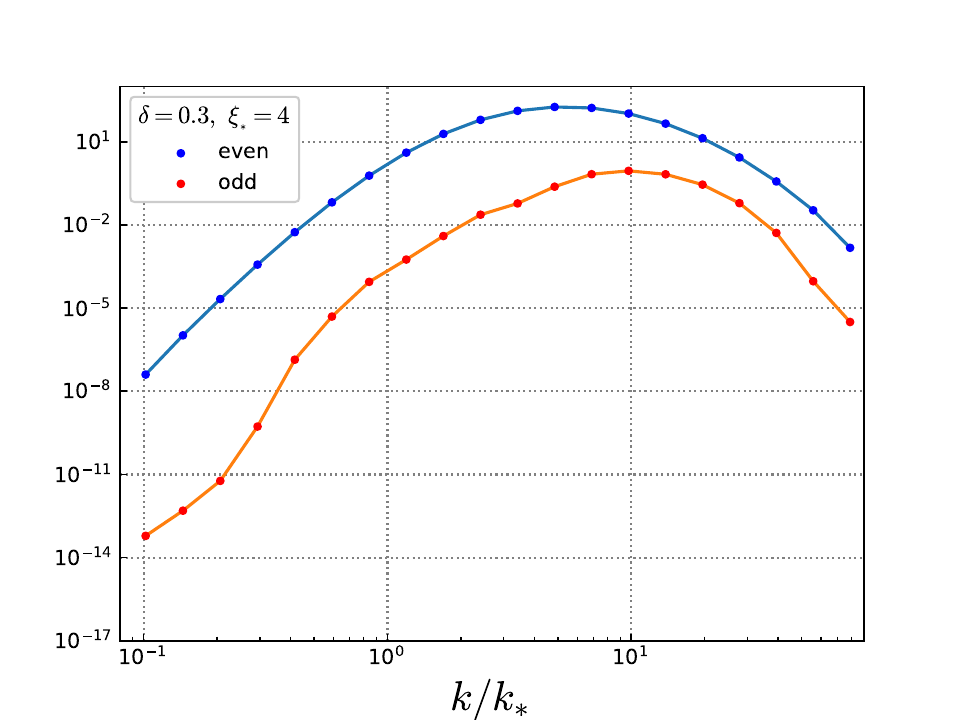}
    \includegraphics[clip, width=0.49\columnwidth]{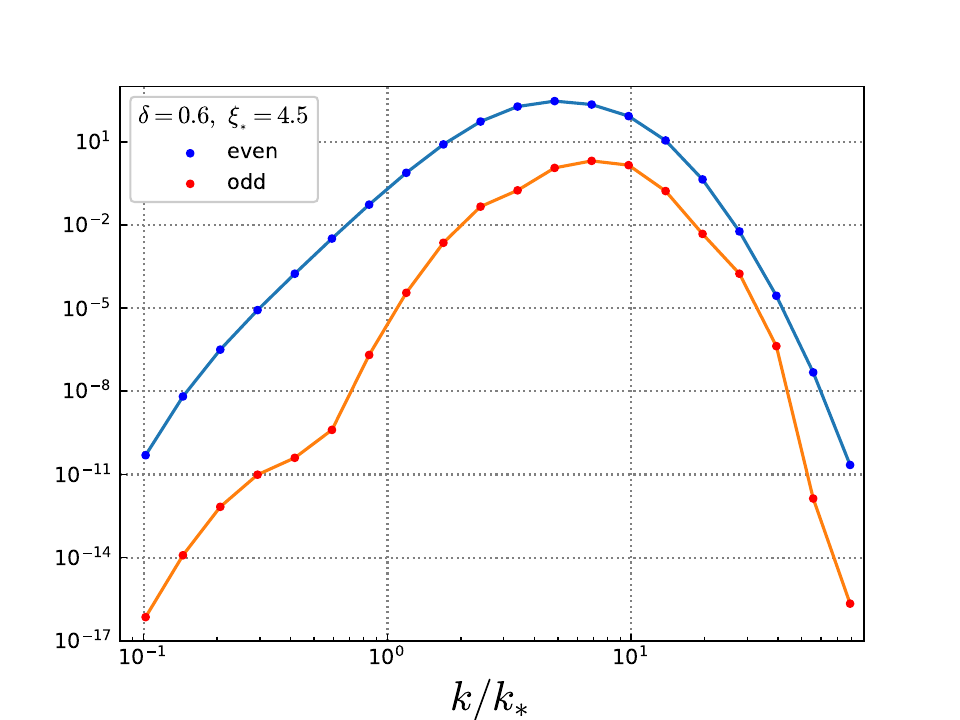}
\caption{
The real part $F_{\zeta, 4}$ (blue line) and the imaginary part $G_{\zeta, 4}$ (orange line) of the normalized trispectrum in the equilateral limit that is averaged over the angular directions defined in Eq.~\eqref{F&G def}. The model parameters are chosen as $\delta=0.3, \xi_*=4$ in the left  and $\delta=0.6, \xi_*=4.5$ in the right panel. 
In the both panels, we use $r_v = 10^{-3.5}$ and $i_{\rm max}=j_{\rm max}=10$.
When the angular average is taken, the parity even part (blue) is typically hundred times larger than the parity odd part (orange). 
}
\label{fig: FG}
\end{center}
\end{figure}

We present the spectral dependence of the averaged trispectum in Figure \ref{fig: FG}.
There we have fixed the model parameters as follows: $\xi_* = 4, \ \delta = 0.3$ in the left panel, $\xi_* = 4.5, \ \delta = 0.6$ in the right panel, and $r_v = 10^{-3.5}, \mathcal{P}_{\zeta, v} = 10^{-9}$ in the both panels.
These parameters are consistent with the constraint on the power spectrum of axion-U(1) with $k_* = 5\times10^{-3} \ \text{Mpc}^{-1}$ for instance \cite{Campeti:2022acx}.%
\footnote{This model can also generate nonzero scalar and tensor bispectra with similar bumped features~\cite{Namba:2015gja}. Regarding their viability, only the cases for $\delta = 0.2$ and $0.5$ have been discussed in the literature, and Refs.~\cite{Namba:2015gja,Shiraishi:2016yun} found that 
the tensor bispectrum for $\delta = 0.5$ could be captured by a LiteBIRD-level CMB B-mode observation. Since a sharper bump due to larger $\delta$ is likely to enhance its measurability, a similar or even better result could be expected in the case for $\delta = 0.6$, which we studied in this paper.
}
As we have discussed in Sec.~\ref{sec:model},
$\delta$ controls the duration over which $\xi (t)$ is significantly large.
A greater $\delta$ shortens the period of tachyonic instability caused by axion.
Accordingly, as shown in Fig.~\ref{fig: FG}, when $\delta =0.6$, the width of the averaged trispectrum is narrower than when $\delta =0.3$.
We also find from this figure that the parity-odd component of the trispectrum peaks at around $k/k_*\sim 10$
and the ratio between the odd and even component is roughly $1\%$ at the peak.

\begin{figure}[htpb]
\begin{center}
    \includegraphics[clip, width=0.49\columnwidth]{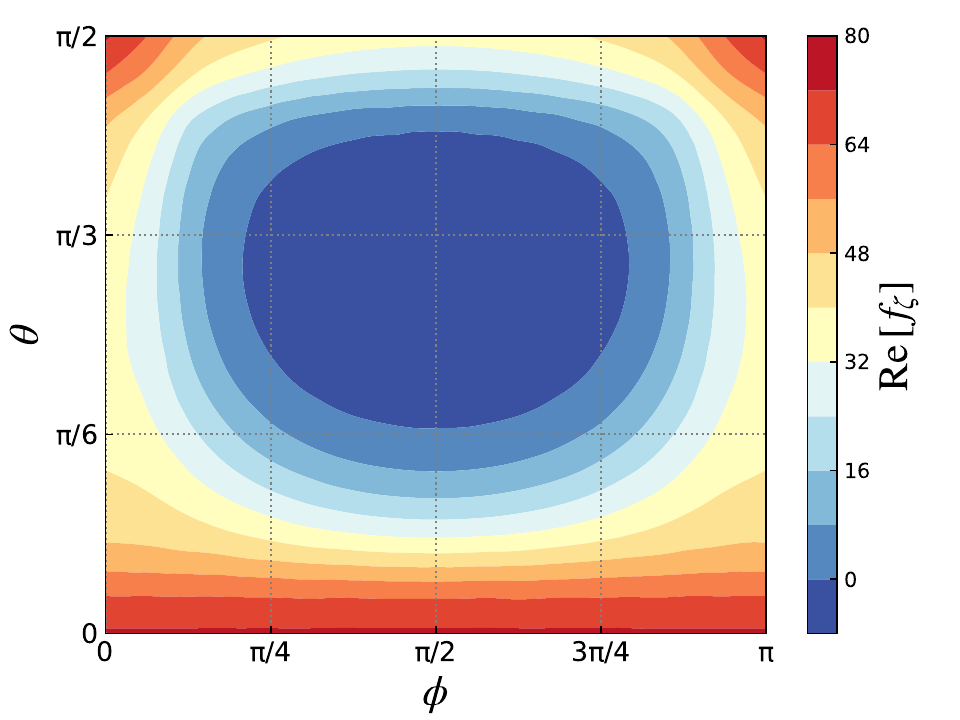}
    \includegraphics[clip, width=0.49\columnwidth]{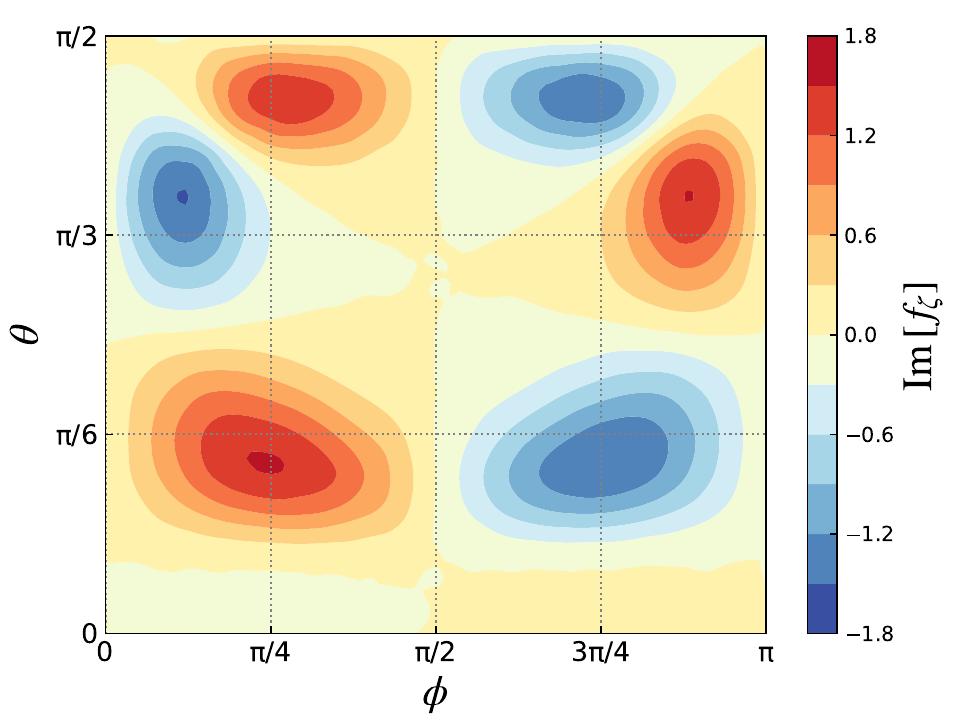}
\caption{
Contour plots of the real part ${\rm Re}[f_{\zeta}^{(\theta,\phi)}]$ (left panel) and the imaginary part ${\rm Im}[f_{\zeta}^{(\theta,\phi)}]$ (right panel) of the normalized trispectrum in the equilateral limit $f_{\zeta}^{(\theta,\phi)}$ introduced in Eq.~\eqref{T and f}.
We numerically compute it based on Eq.~\eqref{eq:f_zeta}.
The color denotes the amplitude of each value.
In this plot, we use $\xi_* = 4.5, \ \delta = 0.6, \ r_v = 10^{-3.5}, k/k_{*}= 10$ and $i_{\rm max}=j_{\rm max}=16$.
}
\label{fig: plot_del6k14_even}
\end{center}
\end{figure}

Next, we investigate the angular dependence of the trispectrum at around the spectral peak, $k/k_*=10$.
In Fig.~\ref{fig: plot_del6k14_even}, we show the amplitude of 
${\rm Re}[f_{\zeta}]$ and ${\rm Im}[f_{\zeta}]$
as a function of $(\phi, \theta)$.
We choose the parameters as $\xi_{*}=4.5, \ \delta=0.6$ and $r_{v}=10^{-3.5}$.
In the case of $|\bm{k}_1 + \bm{k}_2|\to 0$, so called collapsed limit that corresponds to $\theta\to \pi /2$, the polarization vector has a planar structure, so the parity-odd signature vanishes. Similarly, the imaginary part vanishes in $\theta = 0$ and $\phi=0,\pi$ as well. Note that the real part of $f_{\zeta}$ becomes negative  around $\pi /4\lesssim \phi \lesssim 3\pi /4$ and $\pi /6\lesssim \theta \lesssim 5\pi /12$, but otherwise it is positive. Moreover, corresponding to the even (odd) parity condition, ${\rm Re}[f_{\zeta}]$ (${\rm Im}[f_{\zeta}]$) becomes symmetric (antisymmetic) with respect to the $\phi = \pi/2$ line. These mathematical features are apparent from Fig.~\ref{fig: plot_del6k14_even}. Here, we again confirm that the parity-even part ${\rm Re}[f_{\zeta}]$ basically dominates the trispectrum signal.

\begin{figure}[htpb]
\begin{center}
    \includegraphics[clip, width=0.49\columnwidth]{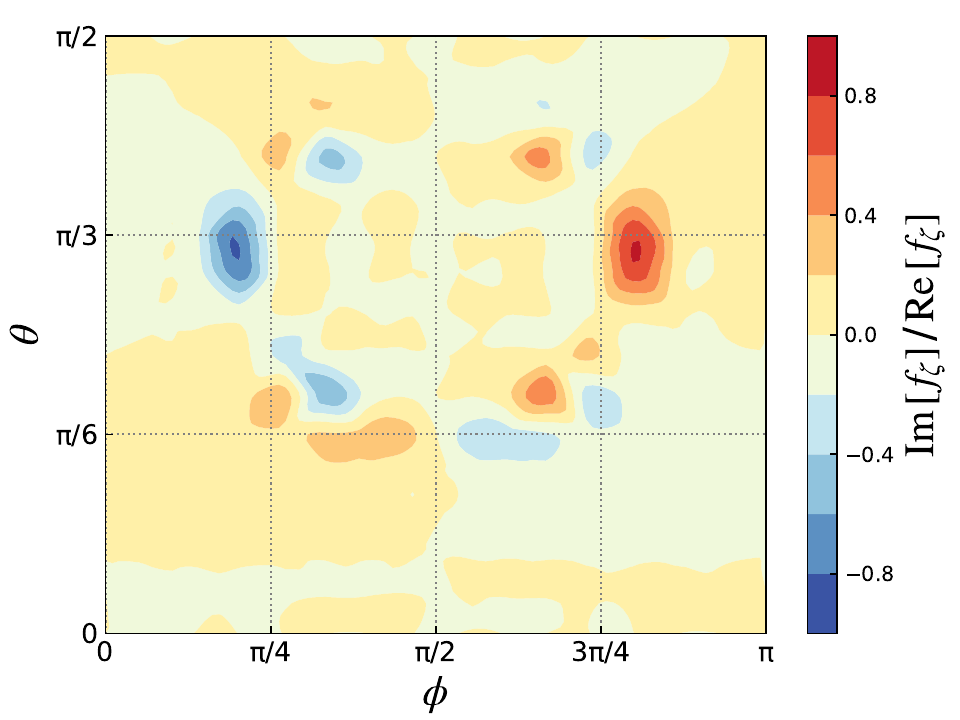}
\caption{
Contour plot of the ratio between the odd and even parity part of the equilateral trispectrum, ${\rm Im}[f_{\zeta}^{(\theta,\phi)}]/ {\rm Re}[f_{\zeta}^{(\theta,\phi)}]$.
In other words, the right panel of Fig.~\ref{fig: plot_del6k14_even} is divided by the left panel.
The parameters are the same as Fig.~\ref{fig: plot_del6k14_even}.
We see that this ratio is of the order of unity in small regions highlighted by the red and blue color.
}
\label{fig: plot_del6k14_ratio}
\end{center}
\end{figure}

In Fig.~\ref{fig: plot_del6k14_ratio}, we plot the fraction of ${\rm Im}[f_{\zeta}] / {\rm Re}[f_{\zeta}]$ as a function of $(\phi, \theta)$, and evaluate the relative size of the parity-odd part. Note that this fraction becomes diverging on the zero-crossing line of the parity-even part ${\rm Re}[f_{\zeta}]$.%
\footnote{In Fig.~\ref{fig: plot_del6k14_ratio}, we have used a cutoff value of order unity to avoid the computational divergences. Nonetheless, such areas are not much resolved in Fig.~\ref{fig: plot_del6k14_ratio} due to a lack of number of bins.}
We here find that the parity-odd part still has $\gtrsim 10\%$ signal of the parity-even one in some specific parameter regions. The $10\%$ fraction is larger than the result from a model of axion as an inflaton analyzed in Ref.~\cite{Niu:2022fki}.
The reason is that, in contrast with the inflaton model, in our spectator model $\xi$ is not directly connected to the curvature perturbation and dynamically evolves in time, making the value of $\xi_{*}$ larger than that in Ref.~\cite{Niu:2022fki}.%
\footnote{
In Ref.~\cite{Niu:2022fki}, $\xi\leq 2.4$ are adopted in the equilateral-limit analysis.
}

\begin{figure}[htpb]
\begin{center}
    \includegraphics[clip, width=0.32\columnwidth]{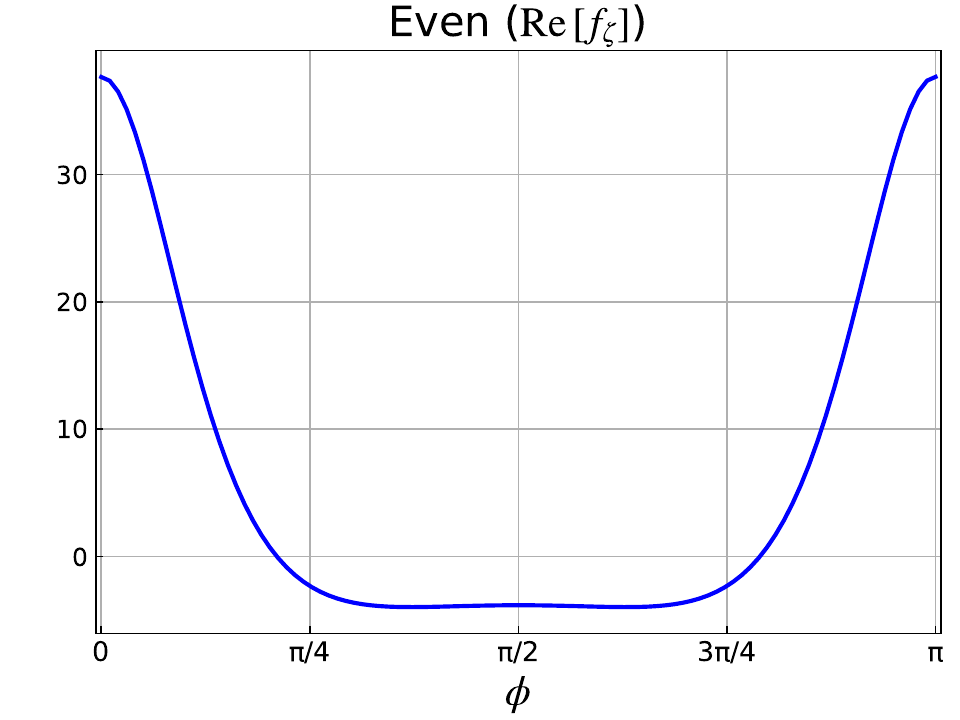}
    \includegraphics[clip, width=0.32\columnwidth]{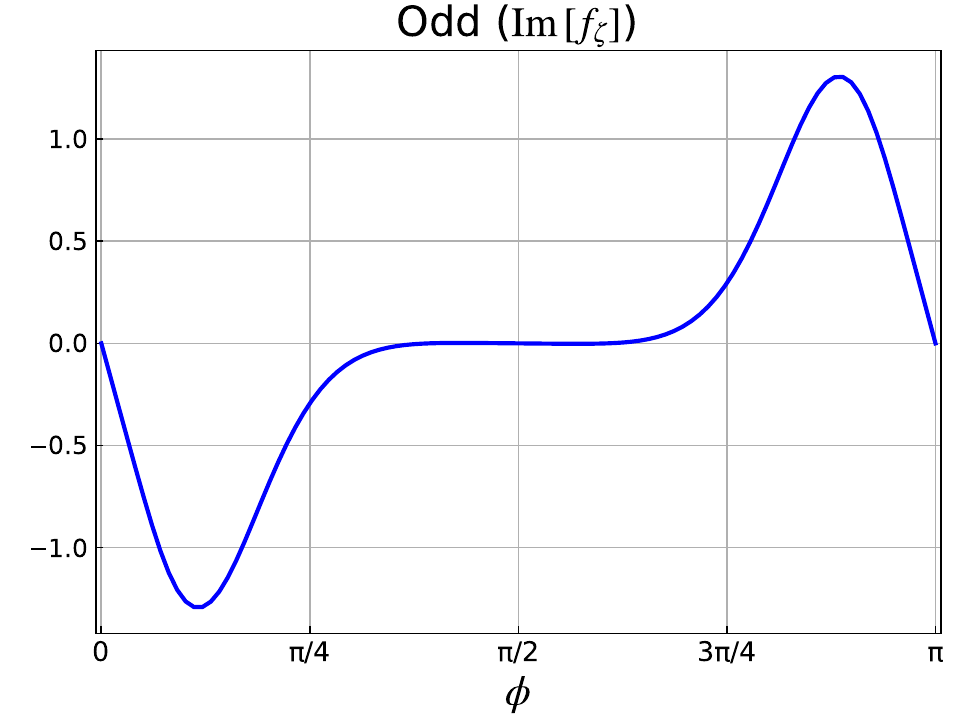}
    \includegraphics[clip, width=0.32\columnwidth]{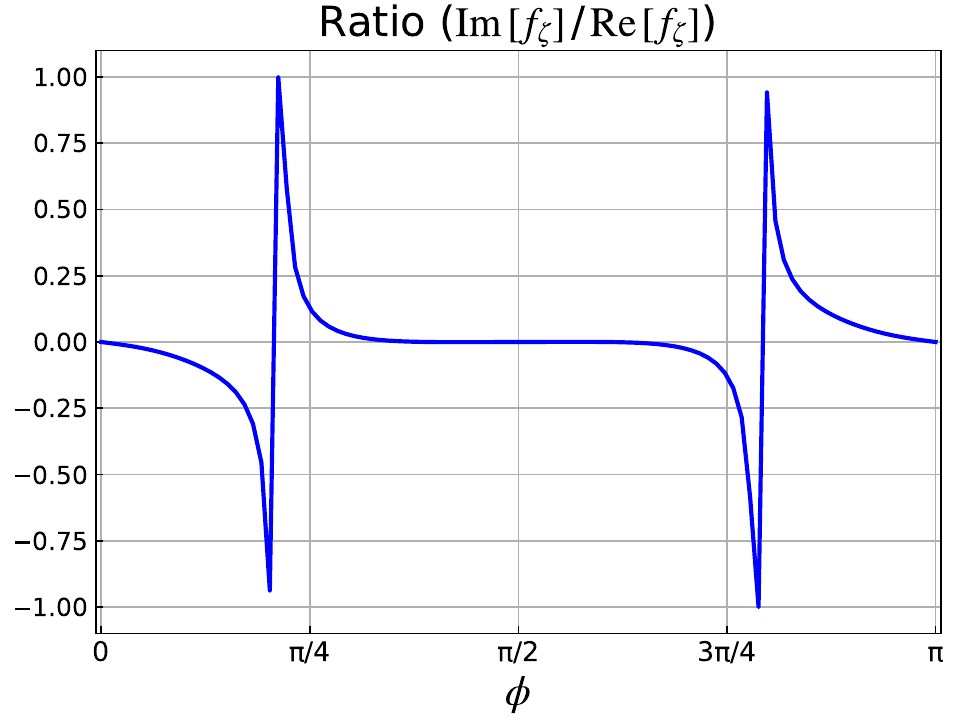}
\caption{
The even part ${\rm Re}[f_{\zeta}^{(\theta,\phi)}]$ (left panel), 
odd part ${\rm Im}[f_{\zeta}^{(\theta,\phi)}]$ (center panel),
and ratio ${\rm Im}[f_{\zeta}^{(\theta,\phi)}]/ {\rm Re}[f_{\zeta}^{(\theta,\phi)}]$ (right panel) shown as functions of $\phi$ by fixing $\theta=\pi /3$. In other words, they are the cross-sections between the plane of $\theta=\pi /3$ and the previous contour plots in Fig.~\ref{fig: plot_del6k14_even} and \ref{fig: plot_del6k14_ratio}.
The parameters are the same as Fig.~\ref{fig: plot_del6k14_even}.
}
\label{fig: plot_del6k14_theta_ratio}
\end{center}
\end{figure}

In Fig.~\ref{fig: plot_del6k14_theta_ratio},
we show $f_{\zeta}^{(\theta,\phi)}$ as a function of $\phi$ with $\theta=\pi/3$.
The parity-even component exhibits zero-crossings at around $\phi \simeq \pi /4$ and $3\pi/4$, and the ratio diverges at the same points.
The parity-odd component peaks at $\phi\simeq \pi/8$ and $7\pi/8$, and the ratio between
the odd and even component reaches $\mathcal{O}(10\%)$ there.
However, again, we do not find any point where the odd component surpasses the even one, except for the close vicinity of the zero-crossing region.

\subsection{Quasi-equilateral shape}

In the equilateral limit, the parity-even part is dominant over the parity-odd part. However, the equilateral configuration is just one of the numerous possible configurations.
In this section, we examine another configuration which is slightly deviated from the exact equilateral one.

\begin{figure}[htpb]
\begin{center}
    \includegraphics[clip, width=0.80\columnwidth]{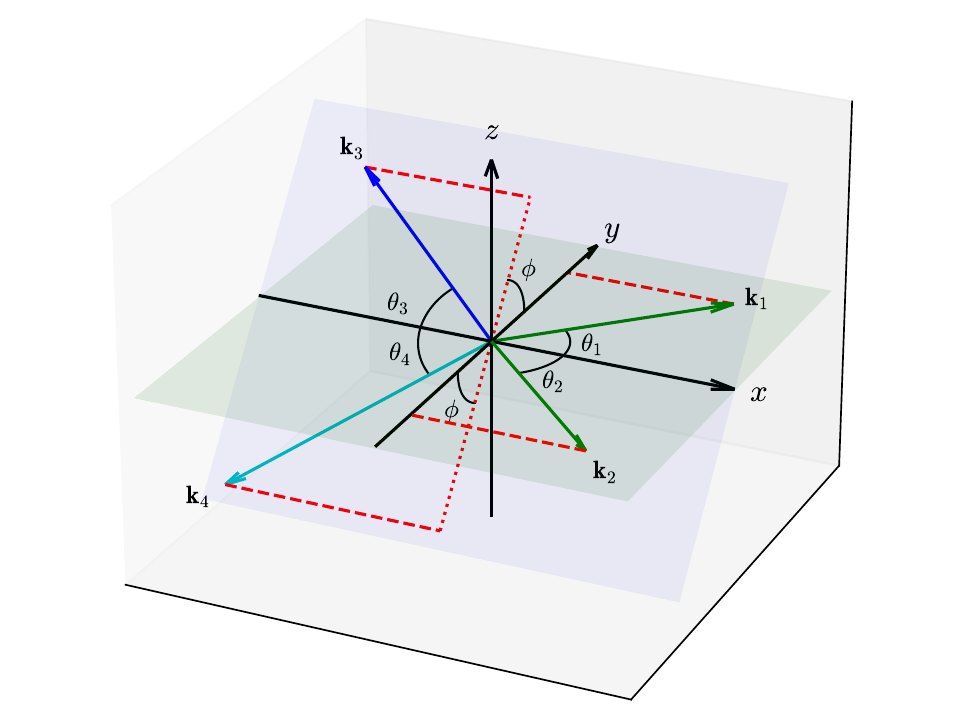}
\caption{
Our momentum configuration in the quasi-equilateral shape.
In the same way as the equilateral configuration in Fig.~\ref{fig: equilateral}, $\bm{k}_1$ and $\bm{k}_2$ are in the $x\text{-}y$ plane (green plane), and $\bm{k}_3$ and $\bm{k}_4$ lie in another plane at an angle of $\phi$ to the $x\text{-}y$ plane around the $x$-axis (blue plane). However, the fourth momentum is longer than the other momenta, $k_4>k_1=k_2=k_3$.
To keep the momentum conservation, $\bm k_1+\bm k_2+\bm k_3+\bm k_4= \bm 0$, the angle to the $x$-axis is generalized from the common value $\theta$ to individually different ones $\theta_i\, (i=1,2,3,4)$.}
\label{fig: non_equilateral}
\end{center}
\end{figure}

We take the coordinates of the momenta with a more generic configuration as shown in Fig.~\ref{fig: non_equilateral}:
\begin{align}
    \bm{k}_1 
    &=
    k_1 (\cos \theta_1, \, \sin \theta_1, 0),
    \\
    \bm{k}_2 
    &=
    k_2 (\cos \theta_2, \, -\sin \theta_2, 0),
    \\
    \bm{k}_3 
    &=
    k_3 (-\cos \theta_3 , \, \sin \theta_3 \cos \phi, \, \sin \theta_3 \sin \phi), 
    \\
    \bm{k}_4
    &=
    k_4 (-\cos \theta_4 , \, -\sin \theta_4 \cos \phi, \, -\sin \theta_4 \sin \phi).
\end{align}
Since two pairs of momenta lie in the same planes as before, we do not need to re-define the angle $\phi$, but we need to generalize $\theta$ to $\theta_i \ (i=1,2,3,4)$ for each of momenta to satisfy the momentum conservation, $\bm{k}_1+\bm{k}_2+\bm{k}_3+\bm{k}_4 = \bm{0}$.

The momentum conservation $\bm{k}_1+\bm{k}_2+\bm{k}_3+\bm{k}_4 = \bm{0}$ leads to the following conditions: 
\begin{align}
&k_1\sin\theta_1 = k_2\sin\theta_2 \ , \\
&k_3\sin\theta_3 = k_4\sin\theta_4 \ , \\
&k_1\cos\theta_1 + k_2\cos\theta_2 = k_3\cos\theta_3 + k_4\cos\theta_4 \ .
\end{align}
It is still tough to cover all considerable shapes.
However, we can guess that a large signal from the generation of the sourced scalar mode is located around the shapes of the four momenta with roughly the same length, because the particle production of gauge field happens around the horizon-crossing.
Therefore, we consider a small deviation from equilateral shape; namely, the case where only one momentum has a different magnitude from the other three momenta
\begin{equation}
k_1 = k_2 = k_3 \equiv k, \quad k_4 \neq k \ . \label{eq: odd}
\end{equation}
Substituting the condition \eqref{eq: odd} and defining $m \equiv k_4/k$, the above conditions are reduced to
\begin{align}
&\theta_1 = \theta_2 \equiv \theta \ , \\
&\sin\theta_3 = m\sin\theta_4 \ , \\
&\cos\theta = \dfrac{1}{2}\left( \cos\theta_3 + m\cos\theta_4 \right) \ .
\end{align}
Then,
we can express $\theta_3$ and $\theta_4$ in terms of $\theta$ and $m$ as
\begin{equation}
\cos\theta_3 = \cos\theta + \dfrac{1-m^2}{4\cos\theta} \ , \quad \cos\theta_4 = \dfrac{1}{m}\left( \cos\theta - \dfrac{1-m^2}{4\cos\theta} \right) \ .
\end{equation}
Thus, 
our angular variables are $\{ \theta, \phi, m \}$.
Notice that the values of $\theta$ and $m$ should be taken to satisfy $0 < \cos\theta_{3,4} < 1$.

\begin{figure}[htpb]
\begin{center}
    \includegraphics[clip, width=0.49\columnwidth]{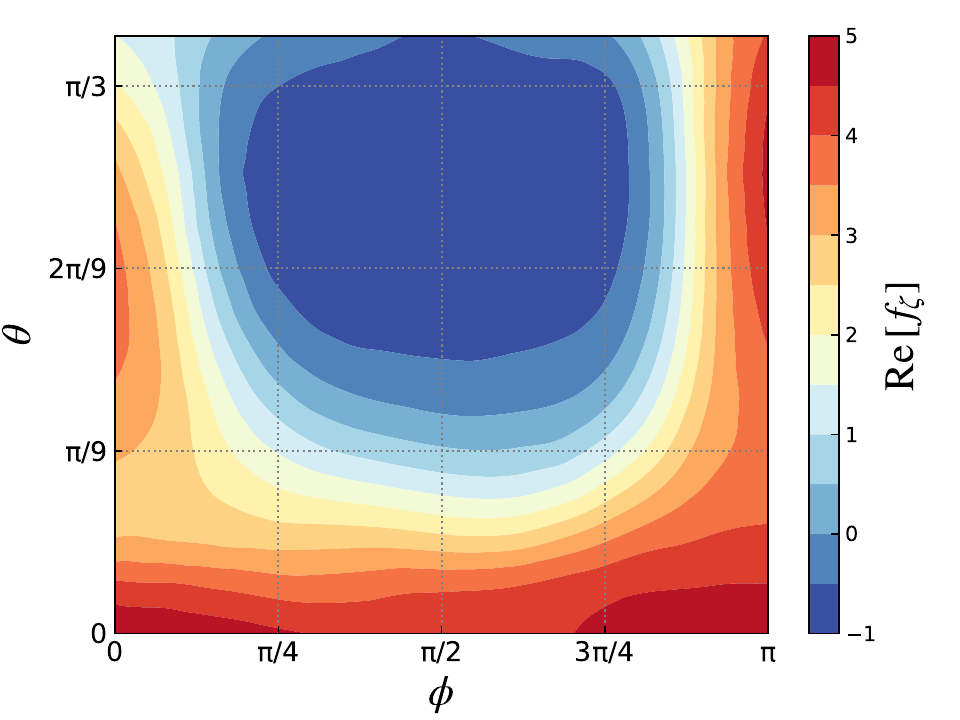}
    \includegraphics[clip, width=0.49\columnwidth]{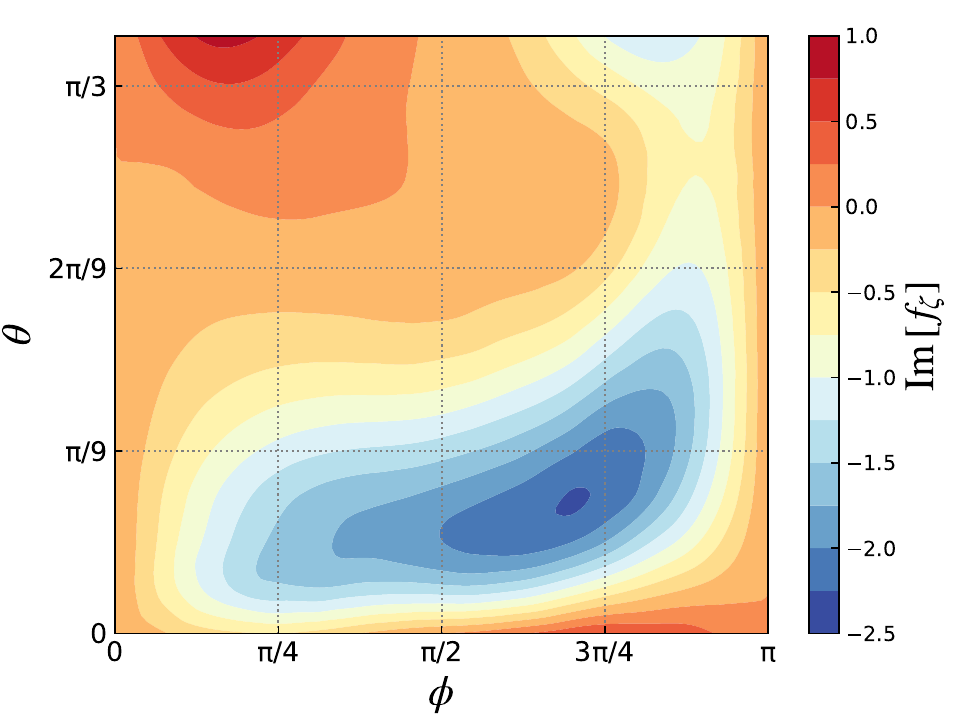}
\caption{
Contour plot of ${\rm Re}[f_{\zeta}^{(\theta,\phi)}]$ (left panel) and ${\rm Im}[f_{\zeta}^{(\theta,\phi)}]$ (right panel) in the quasi-equilateral case.
The parameters are the same as Fig.~\ref{fig: plot_del6k14_even},
except for $m\equiv k_4/k=1.3$.
}
\label{fig: plot_non_del6k14_even}
\end{center}
\end{figure}

\begin{figure}[htpb]
\begin{center}
    \includegraphics[clip, width=0.49\columnwidth]{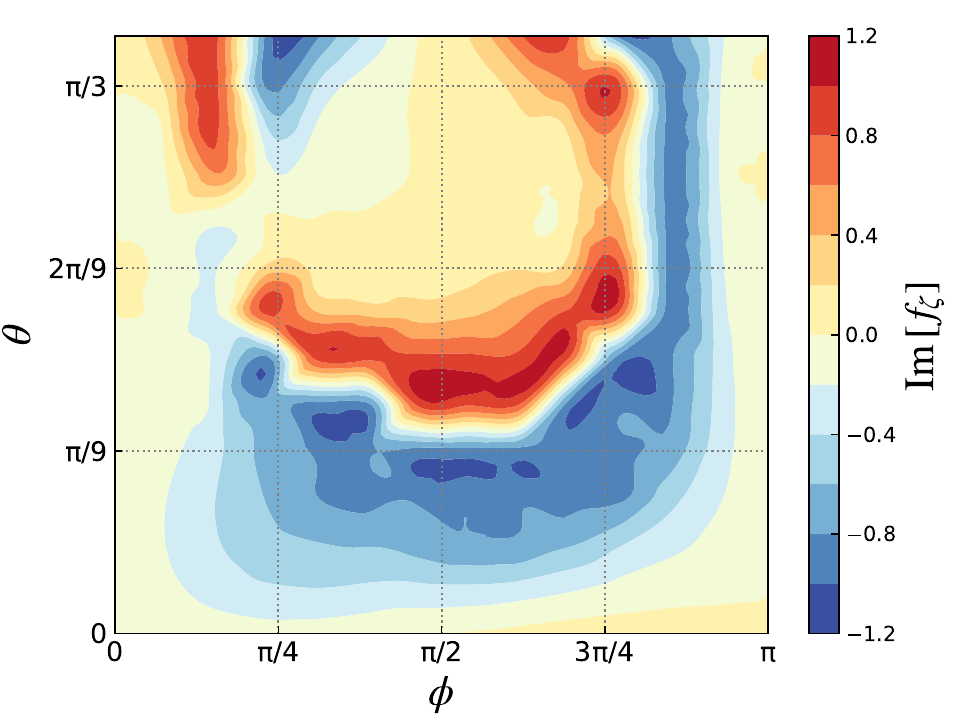}
\caption{
Contour plot of ${\rm Im}[f_{\zeta}^{(\theta,\phi)}]/ {\rm Re}[f_{\zeta}^{(\theta,\phi)}]$ in the quasi-equilateral case.
We set the threshold at $\sim \pm 1$, 
because the ratio diverges when the real part ${\rm Re}[f_{\zeta}]$ crosses zero.
In this plot, we use $\xi_* = 4.5, \ \delta = 0.6, \ r_v = 10^{-3.5}, \ m=1.3$ and $i_{\rm max}=j_{\rm max}=16$.
We see that the ratio becomes of the order of unity in larger regions than the equilateral case.
}
\label{fig: plot_non_del6k14_ratio}
\end{center}
\end{figure}

We numerically compute the dimensionless trispectrum $f_{\zeta}^{(\theta,\phi)}$ in the same way as the equilateral limit case.
In Fig.~\ref{fig: plot_non_del6k14_even}, we show the amplitude of 
${\rm Re}[f_{\zeta}]$ and ${\rm Im}[f_{\zeta}]$
as a function of $(\phi, \theta)$.
We choose the momenta around the inflection point $k/k_{*}= 10$
and parameters are $\xi_{*}=4.5, \ \delta=0.6, \ r_{v}=10^{-3.5},$ and $m=1.3$.
Regarding the even part, its magnitude becomes smaller in comparison with the equilateral case because 
the contribution from the $\bm{k}_4$ mode is slightly off peak and the amplitude becomes smaller than the others.
As a result, we can observe that, in this quasi-equilateral configuration, the real and imaginary parts are of the same order of magnitude
contrary to the case of the equilateral shape.
It can be clearly seen in Fig.~\ref{fig: plot_non_del6k14_ratio} that shows their ratio.
In the darkest red and blue regions of Fig.~\ref{fig: plot_non_del6k14_ratio},  
the absolute value of the ratio, $|{\rm Im}[f_{\zeta}^{(\theta,\phi)}]/ {\rm Re}[f_{\zeta}^{(\theta,\phi)}]|$,
exceeds unity.
Comparing Fig.~\ref{fig: plot_del6k14_even} with Fig.~\ref{fig: plot_non_del6k14_even}, one also notices that 
the real part ${\rm Re} [f_{\zeta}]$ (the imaginary part ${\rm Im} [f_{\zeta}]$)
is no longer symmetric (antisymmetric) with respect to $\phi = \pi/2$ in the quasi-equilateral case.

\begin{figure}[htpb]
\begin{center}
    \includegraphics[clip, width=0.32\columnwidth]{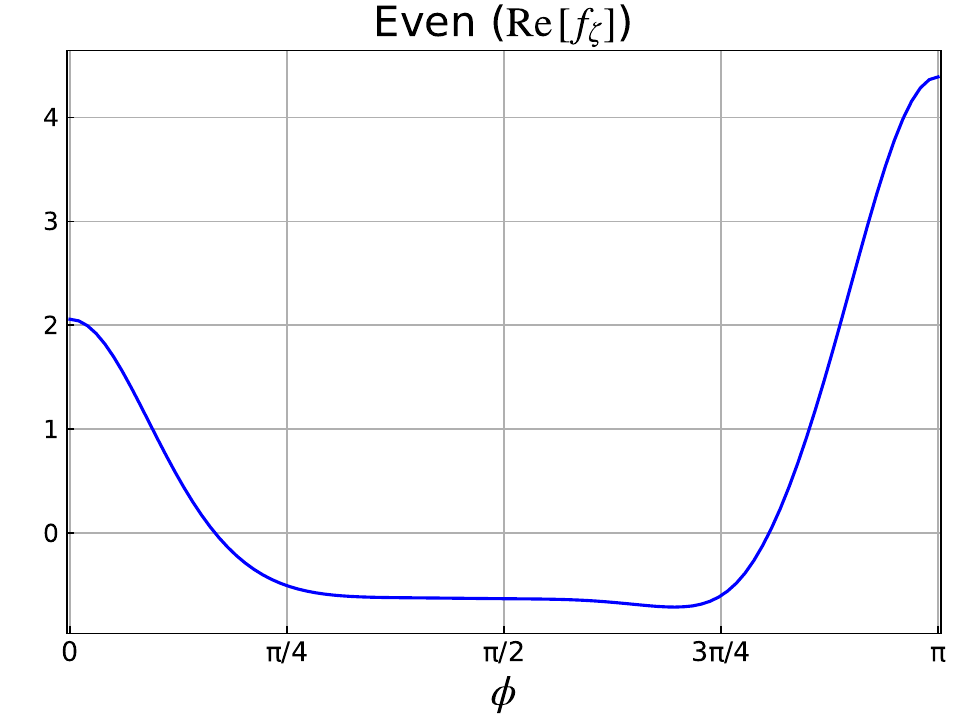}
    \includegraphics[clip, width=0.32\columnwidth]{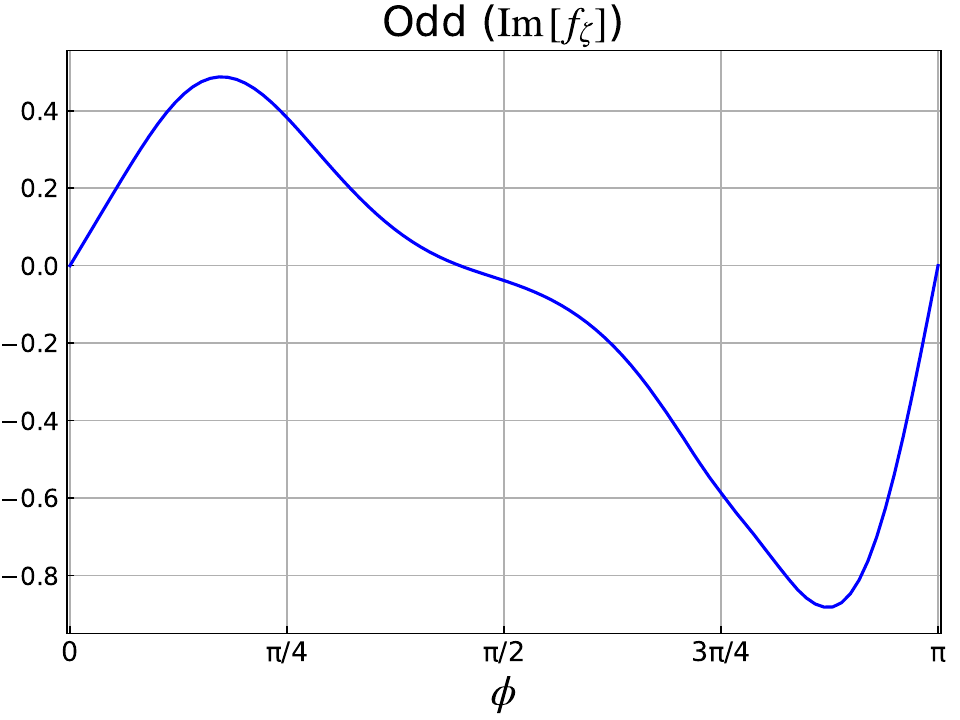}
    \includegraphics[clip, width=0.32\columnwidth]{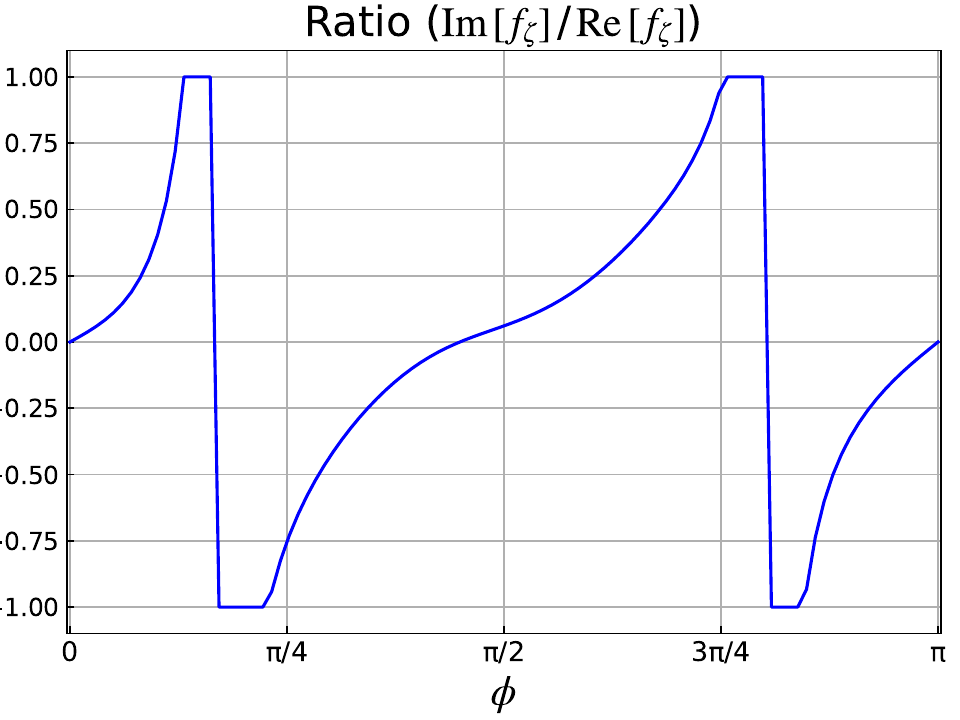}
\caption{
Plot of dimensionless factor of trispectrum \eqref{eq:f_zeta} as a function of $\phi$ in the quasi-equilateral case.
These are even part ${\rm Re}[f_{\zeta}^{(\theta,\phi)}]$ (left panel), 
odd part ${\rm Im}[f_{\zeta}^{(\theta,\phi)}]$ (central panel),
and ratio ${\rm Im}[f_{\zeta}^{(\theta,\phi)}]/ {\rm Re}[f_{\zeta}^{(\theta,\phi)}]$ (right panel) respectively.
We fixed it as $\theta=\pi /3$.
In this plot, we use $\xi_* = 4.5, \ \delta = 0.6, \ r_v = 10^{-3.5}$.
}
\label{fig: plot_non_del6k14_theta_ratio}
\end{center}
\end{figure}

In Fig.~\ref{fig: plot_non_del6k14_theta_ratio},
we present the real and imaginary parts of $f_{\zeta}^{(\theta,\phi)}$ and their ratio as a function of $\phi$ by fixing $\theta=\pi/3$.
The parity-even signal crosses zero at around $\phi \simeq \pi /5,4\pi/5$, and the ratio diverges at the same points.
Contrary to the equilateral shape case, however, the absolute ratio $|{\rm Im}[f_{\zeta}^{(\theta,\phi)}]/ {\rm Re}[f_{\zeta}^{(\theta,\phi)}]|$
exceeds unity even in regions away from the zero-crossing points.
For example, $\phi=3\pi/4$ is close to the negative maximum point of the even component, while the ratio is unity.%
\footnote{
A similar amplification of the parity-odd component in non-equilateral configurations is confirmed also at the bispectrum level \cite{Shiraishi:2012sn}.}
It is also interesting to note that the point where the even component crosses zero and the odd component peaks are roughly coincident.
These results imply that the angular dependence of the trispectrum 
has non-trivial structures and is important to distinguish the origin of the parity-violating signals.

\section{Discussion and conclusion} \label{sec:conclusion}

In this paper, we have studied the generation of the trispectrum sourced by 
the rolling axion coupled to U$(1)$ gauge field during inflation. 
In our setup, one helicity mode of the gauge field is amplified by the tachyonic instability, and the axion perturbation is sourced by the gauge field at one-loop level through the Chern-Simons coupling.
Then, the inflaton perturbation is amplified by the axion perturbation through gravitational coupling. This model generates sizable equilateral-type non-Gaussianity in the primordial curvature perturbation.

We have first studied the exact equilateral shape of the curvature trispectrum. 
Regarding the 
angular-averaged trispectrum, 
we have found that both parity-even and parity-odd components have peaks at slightly smaller scale than one 
crossing the horizon when the axion pass through the inflection point in the potential,
and the amplitude of parity-odd mode is two orders of magnitude smaller than that of parity-even mode.
We have also studied the angular dependence of the trispectrum near the peak scale.
We have found that the ratio of parity-even to parity-odd mode amplitude 
has sharp peaks at certain angles.
This is because parity-even mode crosses zero,
and the ratio becomes large near that point.
Apart from such zero-crossing points, the ratio of parity-odd to parity-even mode is typically ${\cal O}(10^{-1})$.
Therefore,
the parity-odd mode is non-negligible but not dominant compared to the parity-even mode in the equilateral configuration.

We then have extended the analysis to quasi equilateral configurations, e.g., where only one of the momenta is slightly longer than the other three. In comparison with the exact equilateral case, interestingly,
the ratio of parity-odd to parity-even mode exceeds unity in wider regions of angular momentum variables including the area far from the zero-crossing points of the parity-even mode.
This implies the powerfullness of the quasi-equilateral signals for assessing this model.
To clarify this, and moreover, to test with observed datasets of the CMB or large-scale structure, more comprehensive analysis covering the whole momentum space is necessary, and we leave it to our future work.

Finally, let us compare our results with those obtained in a previous similar but different study~\cite{Niu:2022fki} where the gauge field is directly coupled to an axionic inflaton, and the speed of background axion field was assumed to be constant.
Ref.~\cite{Niu:2022fki} fixed $\theta = \pi /3$ in equilateral form and studied the angular dependence of $\phi$.
The trispectra of ours and in Ref.~\cite{Niu:2022fki} have similar features such as the parity-even mode is an even function with $\phi = \pi/2$, while the parity-odd mode is an odd function. Also, for $\phi = 0,\pi$, the parity-odd mode becomes zero because the momenta vector has a planar structure.
On the other hand, differently from our case, the parity-even mode does not cross zero in Ref.~\cite{Niu:2022fki}; therefore, the ratio never becomes very large.
Moreover, in our results, the ratio of parity-odd to parity-even mode at its maximum is ${\cal O}(10^{-1})$ apart from the zero-crossing regions of the parity-even mode, while in Ref.~\cite{Niu:2022fki}, it is only ${\cal O}(10^{-2})$.
Therefore, even excluding the fact that our results exhibit zero-crossings and we studied the quasi-equilateral shape, we found that the parity-odd signal is larger than what was evaluated in Ref.~\cite{Niu:2022fki}.
The reason is that our spectator model imprints unique scale dependence in correlation functions, and
it allows us to make the value of $\xi_{*}$ larger than the inflaton model in Ref.~\cite{Niu:2022fki} (an upper bound $\xi \leq 2.4$ is used in Ref.~\cite{Niu:2022fki}, while this paper obeyed observational bounds in Ref.~\cite{Campeti:2022acx}). In light of this, our scenario may be more useful on a physical interpretation of the large parity-odd trispectrum signal observed in the BOSS data.

In our model, the trispectrum is most enhanced at around the equilateral limit.
According to our numerical calculations, the parity-even mode becomes one order of magnitude smaller in quasi-equilateral configurations compared to equilateral configurations.
In contrast, the parity-odd mode does not decrease significantly, even in quasi-equilateral configurations.
Such interesting tendencies seem to come from chiral structures of contractions between polarization vectors due to the gauge field. We cannot conclude yet that these are common in any spectator axion-gauge field model, but expect quantitatively similar results even in similar but slightly different scenarios as in Ref.~\cite{Niu:2022fki}.

\acknowledgments

I.\,O. and M.\,S. thank to Eiichiro Komatsu and other members in the Japan Society for the Promotion of Science (JSPS) KAKENHI project (No.~JP20H05859) for fruitful discussions. This work is supported by the JSPS KAKENHI Grant Nos.~JP20H05854 (T.\,F.), ~JP23K03424 (T.\,F.), ~JP23KJ2007 (T.\,M.), ~19K14702 (I.\,O.), ~JP20H05859 (I.\,O. and M.\,S.) and JP23K03390 (M.\,S.).


\appendix

\section{WKB solution}
\label{app: WKB}

The differential equation \eqref{eq: eomA} is unsolvable in a closed form due to the dynamical $\xi(\tau)$. Here we find the approximate solution of $A^+_k$ by relying on the WKB method.
The equation of motion is rewritten as
\begin{equation}
\left[ \partial_\tau^2 + \omega(\tau)^2 \right]A^+_k(\tau) = 0 \ , \qquad \omega(\tau)^2 \equiv k^2 + \dfrac{2k\xi(\tau)}{\tau} \ . \label{eq: eomWKB}
\end{equation}
The WKB approximation is valid when the mode function adiabatically evolves in time: $|\omega'(\tau)| \ll \omega(\tau)^2$.
Since we are interested in a regime where $\omega(\tau)^2$ becomes negative and the tachyonic instability happens, we consider a time window $\tau > \bar{\tau}$ at which $\omega(\bar{\tau})^2 = 0$.
Then, defining $\rho(\tau)^2 \equiv -\omega(\tau)^2$, namely
\begin{equation}
\rho(\tau) \equiv \sqrt{-\dfrac{2k\xi(\tau)}{\tau}-k^2} \ ,
\end{equation}
we obtain a WKB solution
\begin{align}
A^+_k(\tau>\bar{\tau}) \simeq \dfrac{C_1}{\sqrt{\rho(\tau)}}\exp\left[ \int^{\bar{\tau}}_\tau \rho(\tilde{\tau})d\tilde{\tau} \right] + \dfrac{C_2}{\sqrt{\rho(\tau)}}\exp\left[ -\int^{\bar{\tau}}_\tau \rho(\tilde{\tau})d\tilde{\tau} \right]
\end{align}
in the regime $|\omega'(\tau)| \ll \omega(\tau)^2$.
The integration constant $C_{1,2}$ is given by $C_1 = 1/\sqrt{2k}, \ C_2 = -i/\sqrt{2k}$ from a boundary condition in the adiabatic vacuum.
To compute the integration function, we separate the time integral $\int_\tau^{\bar{\tau}}$ into two parts: $\int_0^{\bar{\tau}}+\int_\tau^{0}$, where the first part is a constant term and the second part leads to a time-dependent function on $\tau$.
Using the slow-roll solution for $\xi(\tau)$ \eqref{eq: eomA}, the time integral is performed as
\begin{equation}
\int^0_\tau \rho(\tilde{\tau}) d\tilde{\tau} \simeq
\dfrac{4\xi_*^{1/2}}{1+\delta}\left(\dfrac{\tau}{\tau_*}\right)^{\delta/2}(-k\tau)^{1/2}{_2}F_1\left(\tfrac{1}{2}, \ \tfrac{1+\delta}{4\delta}; \ \tfrac{5\delta+1}{4\delta}; \ -\left(\tfrac{\tau}{\tau_*}\right)^{2\delta}\right) \ . \label{eq: hyp}
\end{equation}
We notice that ${_2}F_1$ becomes unity in the limit $|\tau/\tau_*| \rightarrow 0$ and \eqref{eq: hyp} then corresponds to what has been derived in the previous work \cite{Namba:2015gja}.
Finally, the growing mode of WKB solution is expressed as
\begin{align}
  A^+_k &\simeq \dfrac{N(\xi_*, -k\tau_*, \delta)}{\sqrt{2k}}\left(\dfrac{-k\tau}{2\xi(\tau)}\right)^{1/4}
  \notag\\
  &\quad\times\exp\left[ -\dfrac{4\xi_*^{1/2}}{1+\delta}\left(\dfrac{\tau}{\tau_*}\right)^{\delta/2}(-k\tau)^{1/2}{_2}F_1\left(\tfrac{1}{2}, \ \tfrac{1+\delta}{4\delta}; \ \tfrac{5\delta+1}{4\delta}; \ -\left(\tfrac{\tau}{\tau_*}\right)^{2\delta}\right) \right] \ . \label{eq: hyper}
\end{align}

\begin{figure}[thpb]
\begin{center}
    \includegraphics[clip, width=0.32\columnwidth]{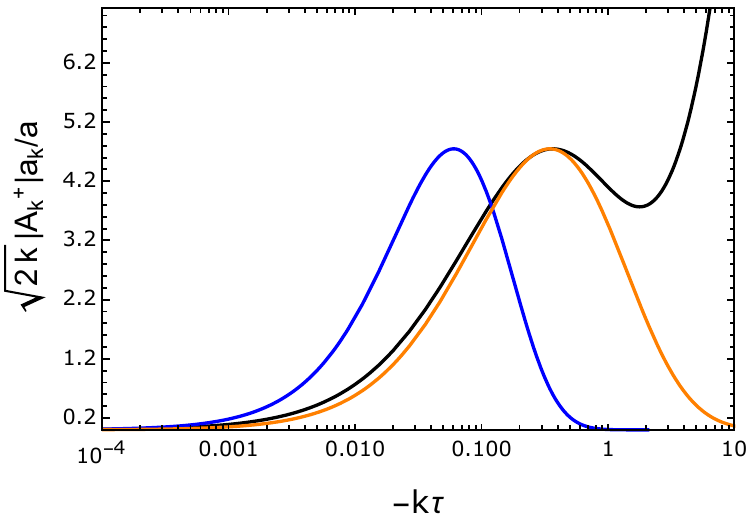}
    \includegraphics[clip, width=0.32\columnwidth]{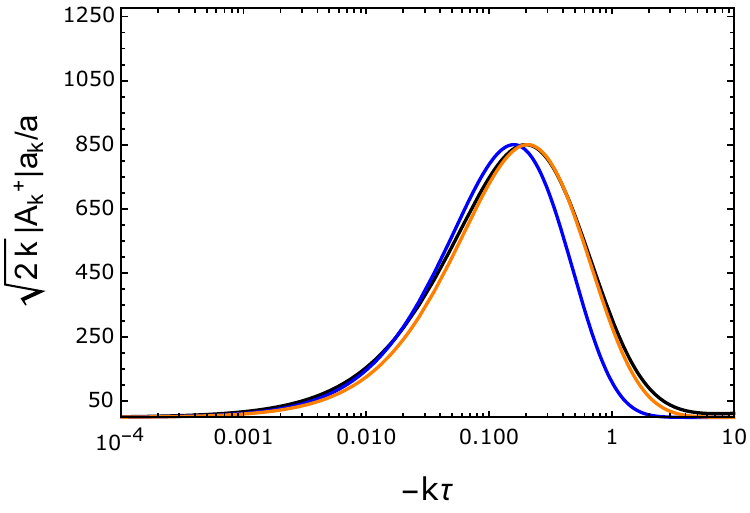}
    \includegraphics[clip, width=0.32\columnwidth]{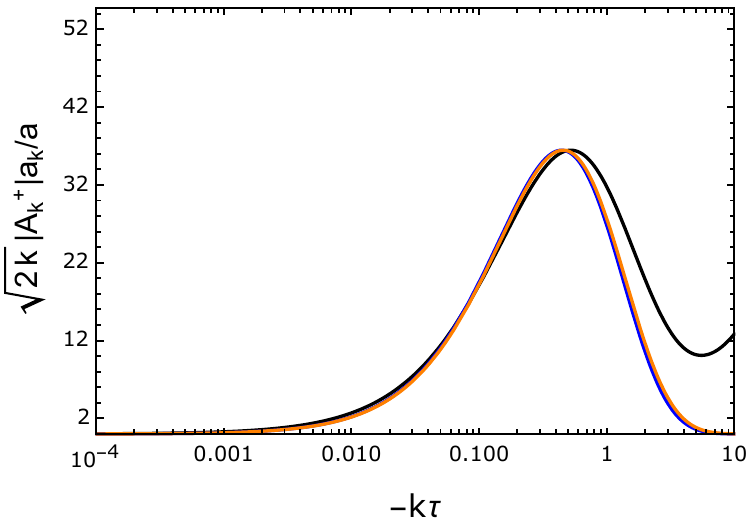}
\caption{
  The time evolution of mode function with $-k\tau_*=10^{-2}$ (left panel), $-k\tau_*=1$ (middle panel) and $-k\tau_*=10^2$ (right panel).
  The horizontal axis is a dimensionless time flowing from the right to the left.
  The black lines denote the exact numerical solutions.
  The orange lines are the WKB approximate solution \eqref{eq: hyper}.
  The blue lines are the solution \eqref{eq: hyper} where ${_2}F_1$ is replaced with $1$.  
  We set the parameter $\xi_*=4$ and $\delta = 0.3$.
}
\label{fig: WKB}
\end{center}
\end{figure}

In Fig.~\ref{fig: WKB}, we depict the time evolution of mode function and compare the numerical solution with the approximate WKB solutions.
The WKB solution \eqref{eq: hyper} is well fitted with the exact solution for the whole momentum modes.
Comparing that, the reduced WKB solution is not good for lower values of $-k\tau_*$ where the approximation $|\tau/\tau_*| \ll 1$ is not so accurate, while it's a well approximation for momentum modes where the amplification mostly happens. 
%

\bibliographystyle{apsrev4-1}
\bibliography{Ref.bib}

\end{document}